\documentclass{aastex63}
\shorttitle{Big image classifications}
\shortauthors{Teimoorinia et al.}
\graphicspath{{./}{figures/}}

\begin{document}

\title{An astronomical image content-based recommendation system using combined deep learning models in a fully unsupervised mode}

\correspondingauthor{Hossen Teimoorinia}
\email{hossen.teimoorinia@nrc-cnrc.gc.ca, hossteim@uvic.ca}

\author{Teimoorinia, Hossen}
\affiliation{NRC Herzberg Astronomy and Astrophysics, 5071 West Saanich Road, Victoria, BC, V9E 2E7, Canada }
 \affiliation{Department of Physics and Astronomy, University of Victoria, Victoria, BC, V8P 5C2, Canada}

\author{Shishehchi, Sara}
\affiliation{Pegasystems Inc. 1 Rogers Street, Cambridge,  MA 02142, USA}

\author{Tazwar, Ahnaf}
\affiliation{Department of Statistics,
Faculty of Science,
University of British Columbia, Vancouver, BC, V6T 1Z4, Canada}

\author{Lin, Ping}
\affiliation{Department of Astronomy and Astrophysics, University of Toronto, Toronto, ON, M5S 3H4, Canada}

\author{Archinuk, Finn}
\affiliation{Department of Physics and Astronomy, University of Victoria, Victoria, BC, V8P 5C2, Canada}

\author{Gwyn, Stephen D. J.}
\affiliation{NRC Herzberg Astronomy and Astrophysics, 5071 West Saanich Road, Victoria, BC, V9E 2E7, Canada }

\author{Kavelaars, J. J.}
\affiliation{NRC Herzberg Astronomy and Astrophysics, 5071 West Saanich Road, Victoria, BC, V9E 2E7, Canada }
 \affiliation{Department of Physics and Astronomy, University of Victoria, Victoria, BC, V8P 5C2, Canada}

\begin{abstract}

We have developed a method that maps large astronomical images onto a two-dimensional map and clusters them. A combination of various state-of-the-art machine learning (ML) algorithms is used to develop a fully unsupervised image quality assessment and clustering system. Our pipeline consists of a data pre-processing step where individual image objects are identified in a large astronomical image and converted to smaller pixel images. This data is then fed to a deep convolutional autoencoder jointly trained with a self-organizing map (SOM). This part can be used as a recommendation system. The resulting output is eventually mapped onto a two-dimensional grid using a second, deep, SOM. We use data taken from ground-based telescopes and, as a case study, compare the system’s ability and performance with the results obtained by supervised methods presented by \cite{Teimoorinia20a}. The availability of target labels in this data allowed a comprehensive performance comparison between our unsupervised and supervised methods.  In addition to image-quality assessments performed in this project, our method can have various other applications. For example, it can help experts label images in a considerably shorter time with minimum human intervention. It can also be used as a content-based recommendation system capable of filtering images based on the desired content.

\end{abstract}

\keywords{Astronomy data analysis - Convolutional neural networks - Neural networks - Astronomy data modeling - Astronomy data visualization }

\section{Introduction}
\label{introduction}

 Every day, large volumes of data are produced at an unprecedented rate in different domains, 
including astronomical imaging. Since 2003 the Canada-France-Hawaii Telescope's 378 mega-pixel MegaCam \citep{Boulade03} has been producing $\sim 734$~MiB images at the rate of 100 or more per observing night. Operational since 2012, the Dark Energy Camera on the Cerro Tololo Inter-American Observatory (CTIO) \citep{2015AJ....150..150F} is made up of 570 megapixels while the Subaru Telescope's 870 mega-pixel Hyper Suprime Camera (HSC) \citep{2018PASJ...70S...1M} produces nearly 2GiB per image each with a higher frame acquisition rate than CFHT MegaCam.  And, the 3,200 megapixel Rubin Observatory LSST Camera \citep{2010SPIE.7735E..0JK} is scheduled to begin operations in 2022 and will acquire exposures at an even higher cadence.  These examples of ever increasing camera size and desire to create a higher and higher frequency sampling of the night sky has created a deluge of astronomical imaging whose quality assessment is and will be a monumentally challenging task when approached in a classical way.

Recent developments and novel technologies help store and access this data; however, processing and extracting useful knowledge from the data is a popular and challenging topic in computing research. Various research areas explore data of different sizes, types, and quality. In modern astronomy, the data sets are enormous, complex, and multi-dimensional. So, manual processing of such data is impossible. For example, an image taken by ground-based telescopes contains a wide array of objects captured under various adverse atmospheric conditions and amidst miscellaneous technical issues. Although classical machine learning (ML) methods have proven successful under certain circumstances and frameworks \citep{Skrutskie06, Ivezic04}, they often fail to explore complex data sets, and naturally they struggle when faced with non-linearity in the data.  Therefore, more sophisticated ML methods are required to process such data.

Advanced machine learning methods have successfully dealt with complex and nonlinear data sets—from ranking tabular data sets \citep[e.g.,][]{Teimoorinia16, Bluck20} and classification and clustering spectral data \citep[e.g.,][]{teimoorinia12,rahmani18} to exploring complex images in search of exciting phenomena \citep[e.g.,][]{Pourrahmani18,Jacobs19,Bottrell19, Teimoorinia20b}.  There are numerous algorithms and methods in ML, each suited to specific problems. These methods can be categorized into different groups; however, supervised and unsupervised methods are two of the major categories with many sub-categories. In supervised learning, the training data set has a known label or output value. So, the goal is to infer a function that maps the properties of input data to the result through a training process. Supervised learning is usually categorized as a classification problem \citep[e.g.,][]{Huertas-Company-19,Ciprijanovic-20,Ferreira-20,Teimoorinia20a} where the labels are discrete values, or regression  methods \citep[e,g.,][]{teimoorinia14,ellison16a} where the labels are continuous values. In unsupervised learning, however, there are no labels, so the goal is to infer structures within the input data.

Supervised methods are conditional methods that excel at optimizing performance in a well-defined task where plenty of labels are available. However, they have limited potential to generalize knowledge beyond the assigned items they are trained on. In other words, supervised methods are excellent at solving specific learning problems, such as distinct and well-defined image classifications. Since labeling processes generally need expert knowledge and are expensive and time-consuming, unsupervised methods can help mediate problems. However, an unsupervised method alone cannot label a set of images. Instead, it can group the images based on similarities or dissimilarities. In this way, images in one group are similar to each other but different from those in another group. In this method, instead of labeling an enormous number of unlabeled images, an expert can manually inspect the separate groups in a considerably shorter time. Then labels can be applied to all the members inside each group. 

There are several unsupervised clustering methods, such as Kmeans \citep{Berkeley67} and self-organizing maps (SOMs) \citep{Kohonen82, KO01}. The goal of these methods is to find similar clusters of data using dissimilarity or distance metrics. The Kmeans algorithm aims to find $\it{K}$ clusters by averaging the data. Each data point is assigned to a cluster, often by calculating a Euclidean distance metric between data points in feature space. A Euclidean metric fails when the dimension of input, such as an image, is very high. An SOM is a type of neural network. This method creates a mapping between high-dimensional feature space and low-dimensional output space—often a two-dimensional lattice. The algorithm aims to preserve the distance between data points so that points that are close together in the input space are mapped to units near each other in the output map. However, these methods are mostly shallow and cannot handle data sets with high dimensionalities, such as images. One solution is to reduce the dimensionality of the input data and then apply clustering methods.

There are different approaches to dimensionality reduction. Principal component analysis (PCA) is a linear method that summarizes the input data into fewer dimensions by creating uncorrelated features while minimizing the total distance between data. However, its linear behavior limits its applicability to non-linear complex data sets. A non-linear and more sophisticated method is an autoencoder \citep[e.g., see][for the connection with PCA]{Plaut-18}. Autoencoders are feed-forward neural networks that map input data onto itself in the output layer, within certain limits to prevent copying data to the output layer. An autoencoder is a bottleneck system in which the middle layer is considered representative (latent) data. Choosing an appropriately small number of neurons in the middle layer maps the high-dimensional input data into a low-dimensional space. This data is, therefore, a very compact representation of the input data. For example, utilizing radio-astronomical data \cite{Ralph19}, one can use autoencoder methods to reduce the dimensionality of the data and then employ the compressed representations to cluster data utilizing SOM methods.

To increase clustering efficiency while decreasing training time, some methods combine two ML techniques. In this case, each ML method's result yields a more meaningful input to the other, contributing to each method's better performance while aiming to minimize a joint loss function. For example, \citep{Xei16} present a deep embedded clustering method that jointly trains a Kmeans algorithm with a deep neural network (DNN). The low-dimensional feature space resulting from the DNN mitigates the curse of dimensionality problems \citep[e.g., see][]{bishop:2006} faced by the Kmeans algorithm when input data has high dimensionality.  \cite{Forest2019} also introduce a deep embedded SOM (DESOM), where an autoencoder is jointly trained with a SOM.

Image labeling processes are essential topics in astronomy. For example, selecting and labeling only high-quality images can give researchers more confidence in their research results. \cite{Teimoorinia20a} (hereafter T20) have shown that using a combined ML technique in a supervised mode dramatically increases accuracy and performance. In fact, this supervised method mimics human expert knowledge in labeling astronomical images. The question is whether a machine can do this task with minimum supervision.

This paper will combine two DESOM algorithms to label and assess the quality of astronomical images. The first DESOM presents useful and compressed information derived from huge images. This work's significant contribution is to present a novel way to use the output from the first DESOM and convert it into a form suitable for input into the second DESOM.  In this way, we make a robust labeling system for an instrument. As a case study, we employ the same data used by T20 and compare the two methods' results. Hence, we can compare a fully unsupervised method with supervised methods where experts provide the labels. Our unsupervised method can be applied to different images taken from ground- or space-based telescopes such as the Hubble Space Telescope. In Sec. \ref{sec:data}, we introduce the data used in this paper. Sec. 3 describes the different deep learning methods used in this paper. Sec. \ref{sec:result} presents the result from the pipeline that we developed and compares them with experts' labels. Sec. \ref{sec:discussion} discusses the results.

\section{Data}
\label{sec:data} 

\begin{figure*}
\centering
\includegraphics[width=14.cm,height=14cm,angle=90]{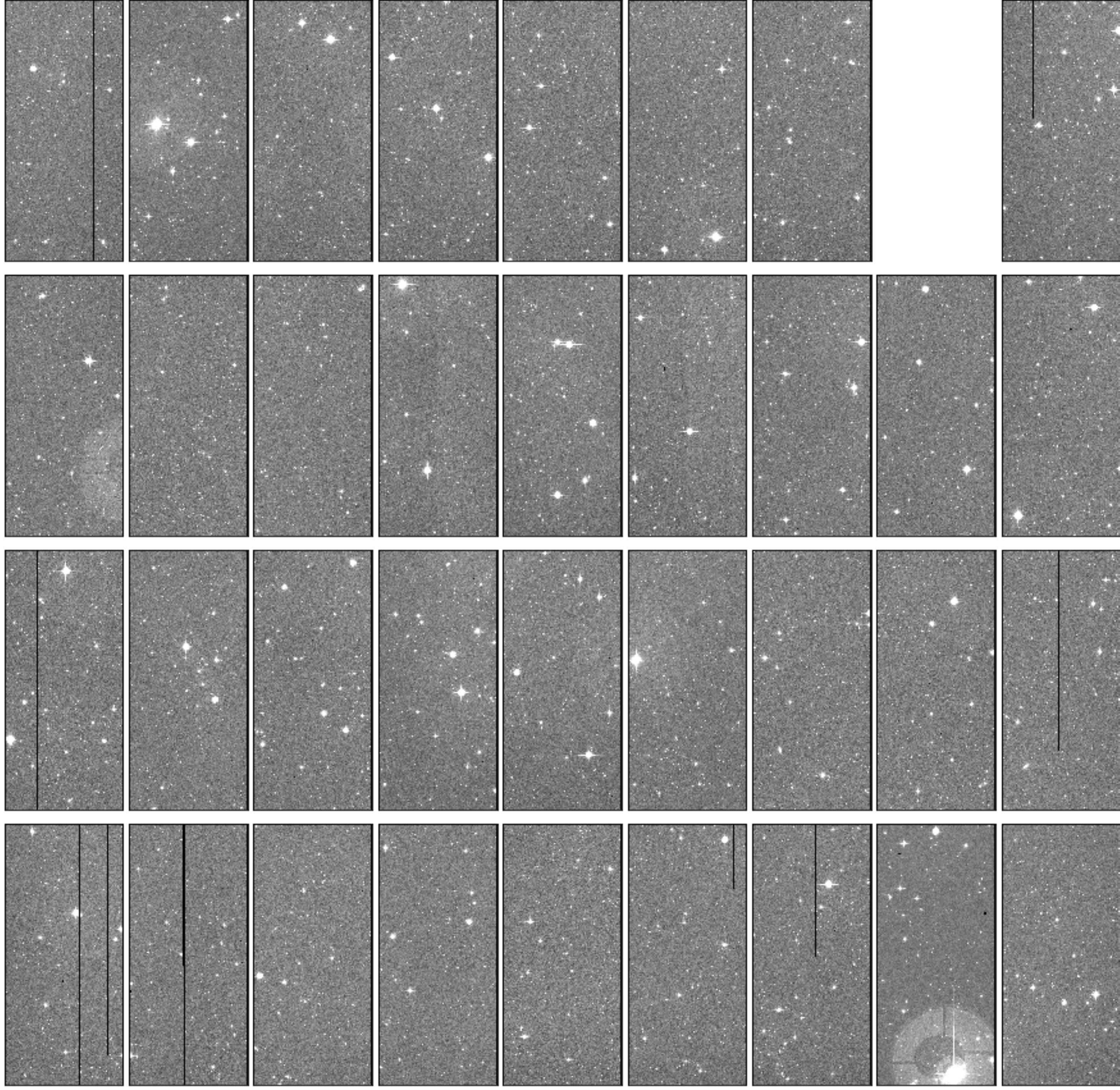}\\
\vspace{.5cm}
\includegraphics[width=7.cm,height=14cm,angle=90]{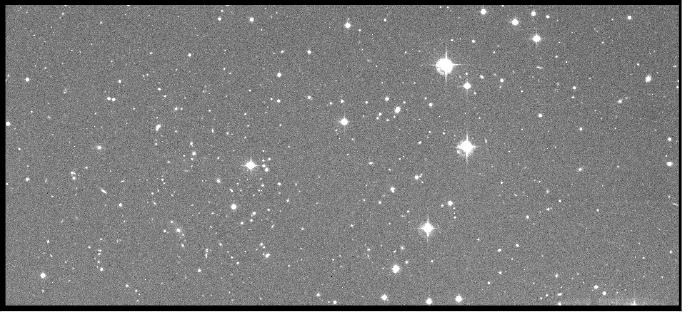}
\caption{The example  of a typical MegaCam exposure consisting of 36 CCDs used by \cite{Teimoorinia20a}. The white section of the image represents a non-functioning CCD caused by sporadic electronic failures. The bottom plot shows one of the 36 CCDs (the upper right one;  $4644\times2112$ pixels) in a larger view.}
\label{fig:36good}
\end{figure*}

\begin{figure*}
\centering

\includegraphics[width=7.cm,height=7cm,angle=0]{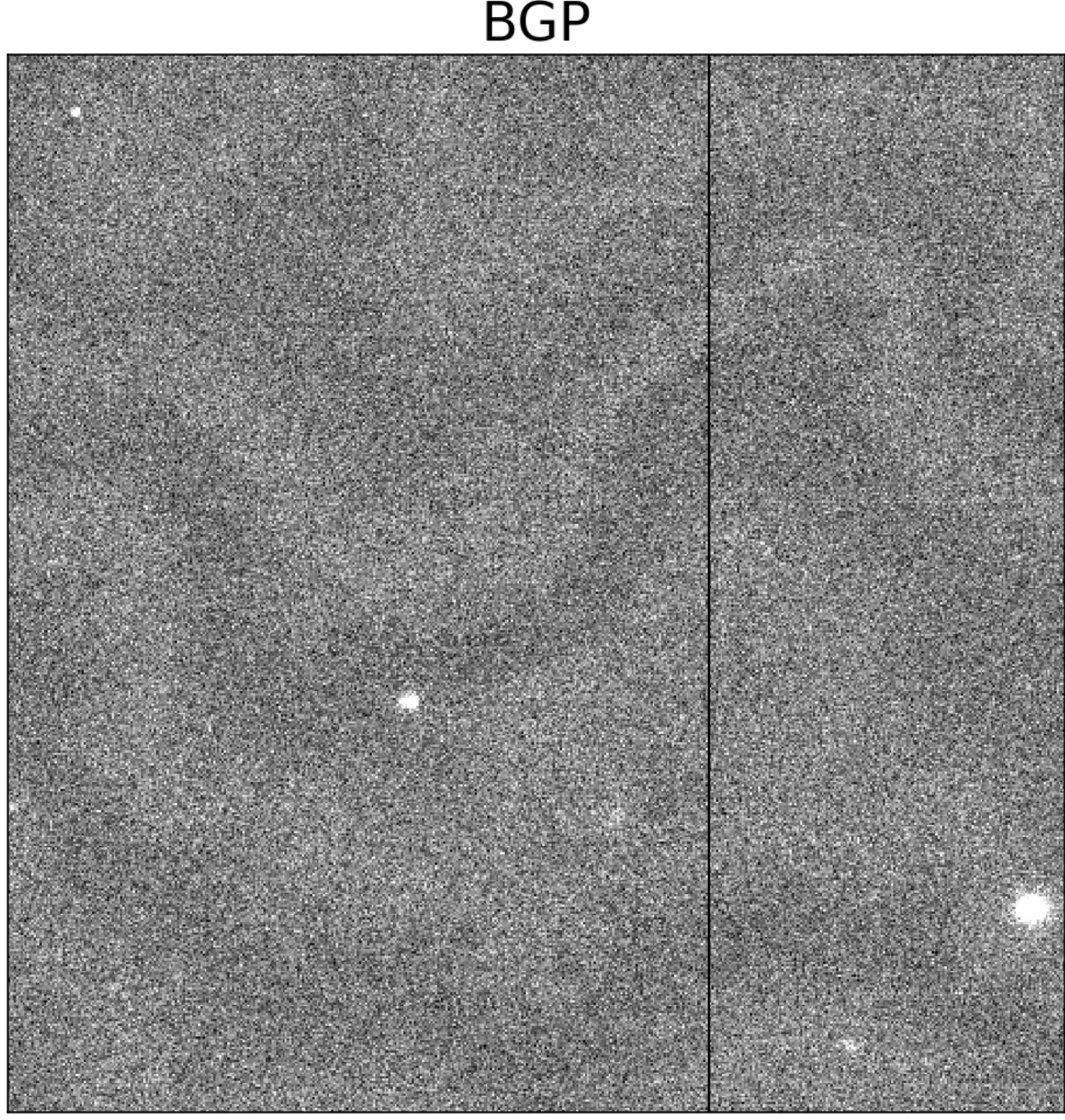}
\includegraphics[width=7.cm,height=7cm,angle=0]{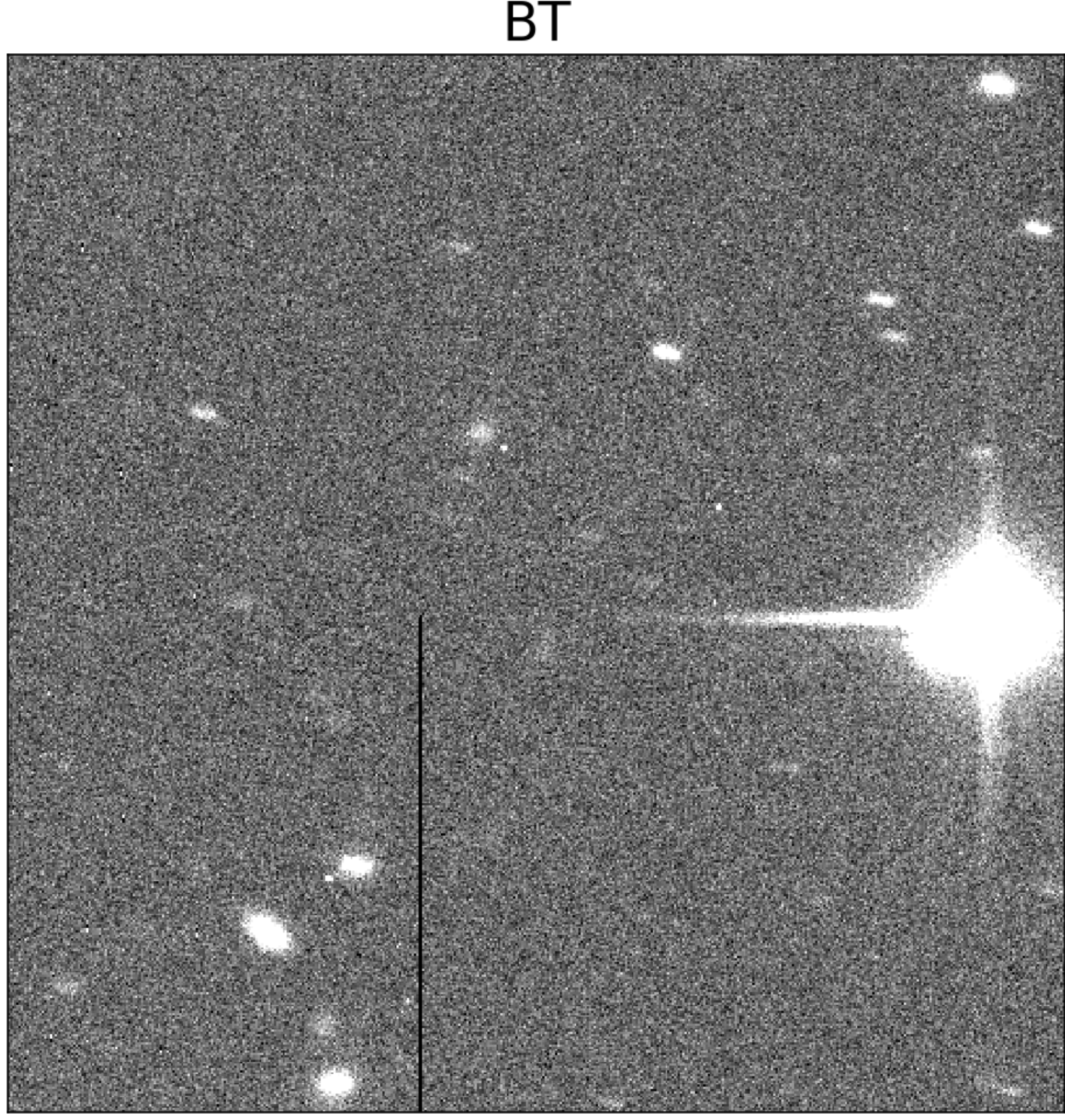}
\includegraphics[width=7.cm,height=7cm,angle=0]{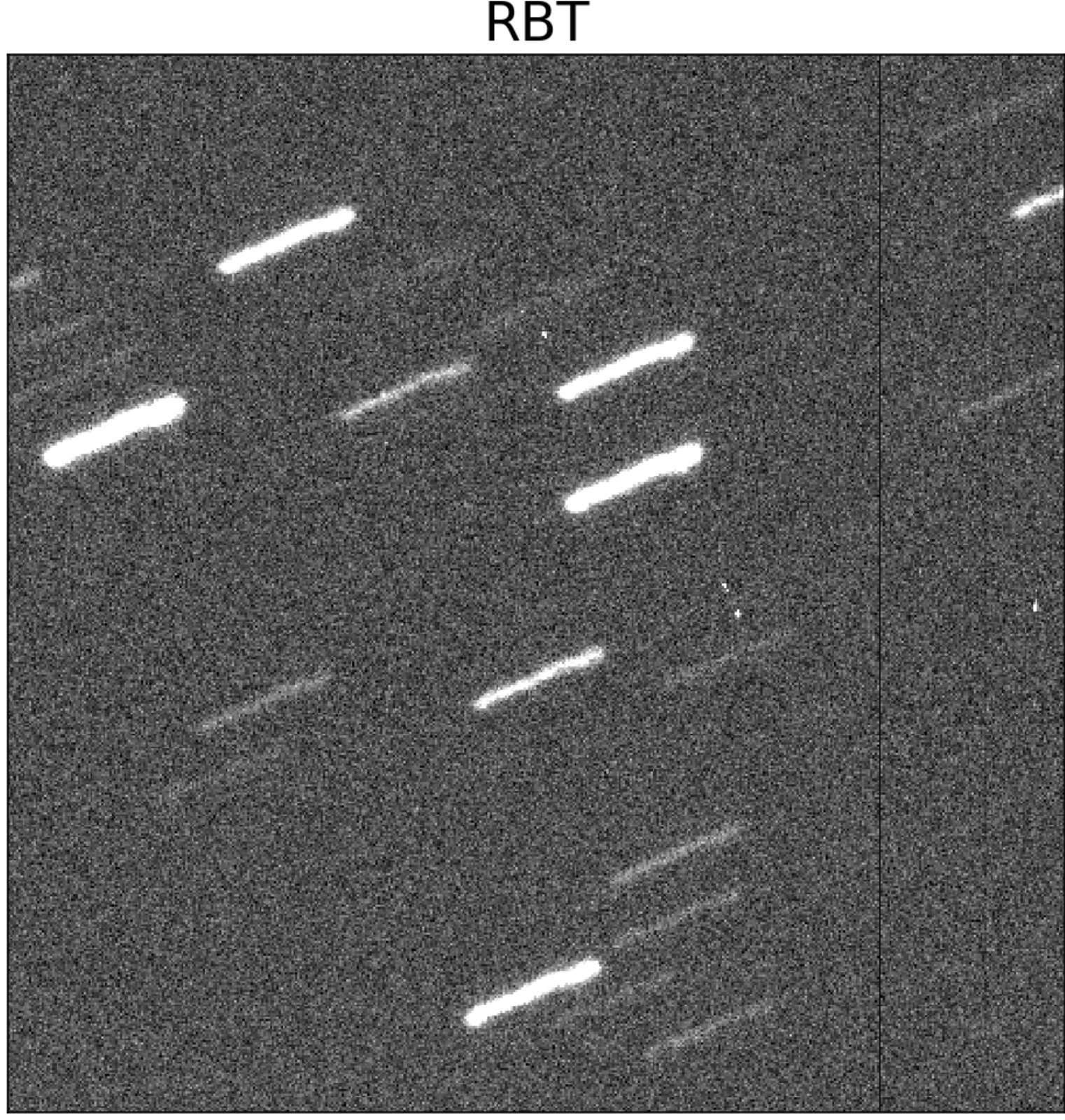}
\includegraphics[width=7.cm,height=7cm,angle=0]{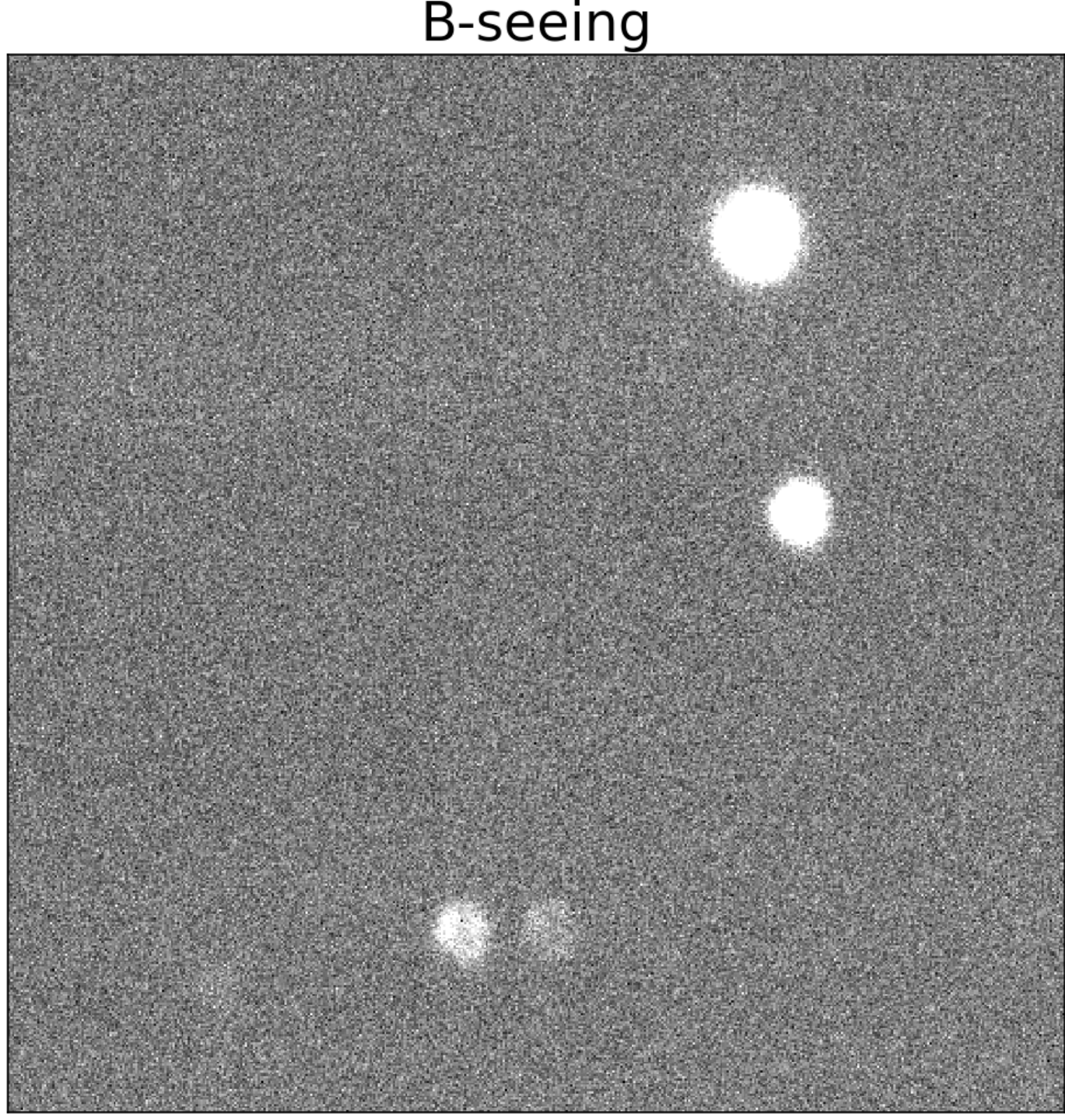}
\includegraphics[width=14.cm,height=7cm,angle=0]{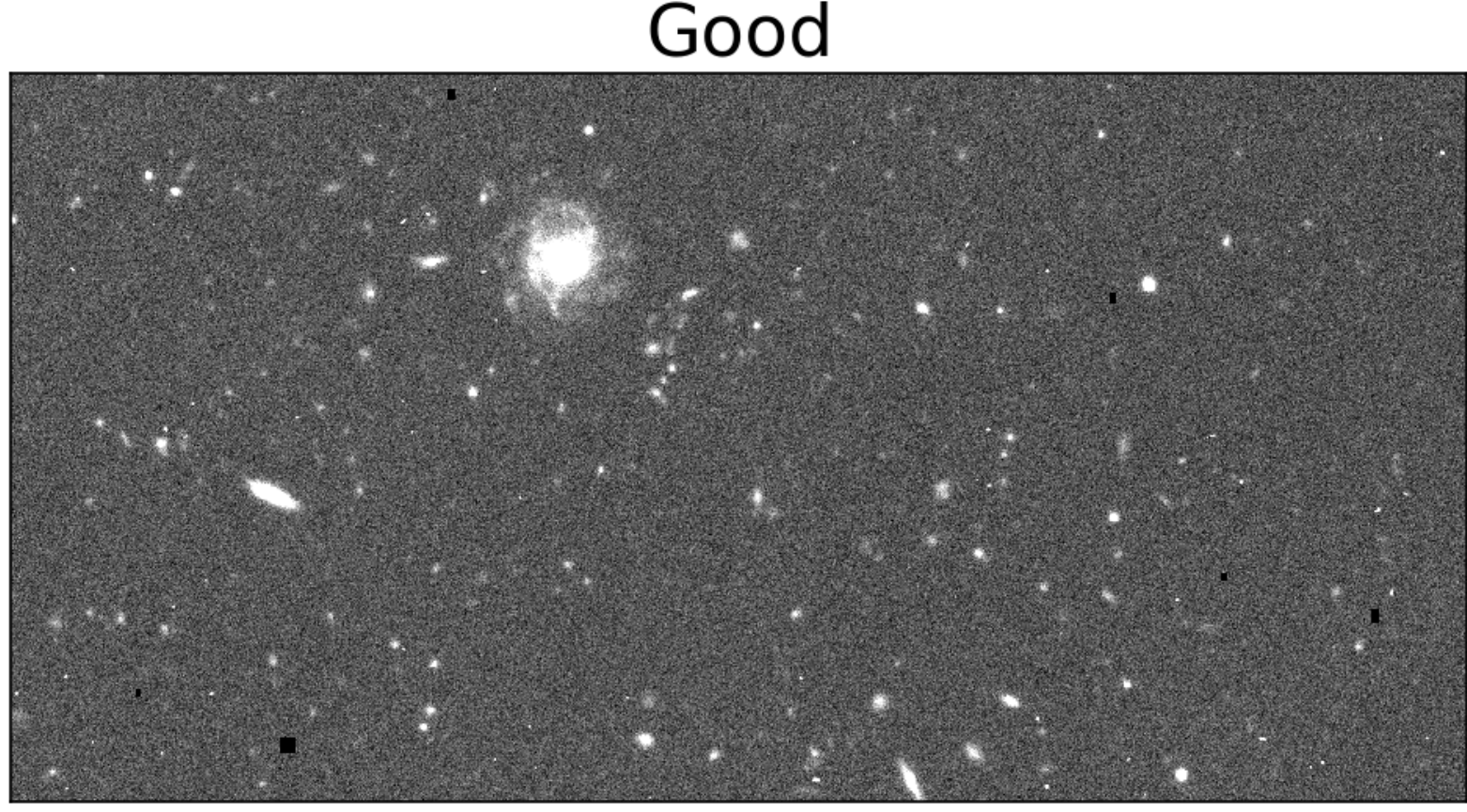}
\caption{The five different targets used by T20 to train a combined deep model. They include images with different problems in the background (BGP), bad tracking (BT), really bad tracking (RBT), bad seeing or bad observational conditions (B-Seeing) and an instance of a “Good” image at the bottom of the figure.}
\label{fig:target}
\end{figure*}

Since we want to compare our unsupervised method with a supervised one, and in order to have a fair comparison, we use the same data explained and used by T20.  They use a set of selected images from the MegaCam instrument mounted on the Canada-France-Hawaii Telescope \citep{Boulade03} to train a combined deep model, classify MegaCam images, and then label them. Ground-based optical telescopes such as MegaCam manifest a wide range of image quality. In the Canadian Astronomy Data Centre, to calibrate and combine astronomical images \citep{gwyn08} the images are first visually inspected to classify them into different image-quality based categories.  In Figure~ \ref{fig:36good}, we show an example of a single exposure image (consisting of 36 CCDs and one non-functional CCD).  T20 use five different classes, including telescope-tracking from slight (BT) to severe (RBT), bad sky conditions (e.g., poor visibility, cloudy conditions, i.e., B-Seeing) and different background problems (BGP), such as unusual background fluctuations, severe object-saturation, and non-astrophysical background patterns. Figure \ref{fig:target} shows a sample of the five classes.

The model's input is taken from a single exposure image that contains 36 (or 40) CCDs/images, with each pixel imaging $\sim 0.184$~arc-second region of the sky.  In this work, the terms `exposure' and `ID' are used interchangeably to mean a MegaCam image with 36 (or 40) CCDs such as that shown in Figure \ref{fig:36good}.  Each CCD has a size of 2048x4612, so a single exposure has $\sim$ 350 million pixels. As a test set, we use the same $\sim70\rm{k}$ exposures and their predicted labels by T20.  In this way, we can compare the supervised and unsupervised models.

\subsection{ A brief review of T20}
T20 present a two-component Machine Learning (ML) based approach for classifying astronomical images.  Using a SOM as the first component, they create a proper and small fraction of the image pixels. The second part, i.e., a deep model, uses the images to determine the observation's quality.

\begin{figure*}
\centering
\includegraphics[width=18cm,height=9cm,angle=0]{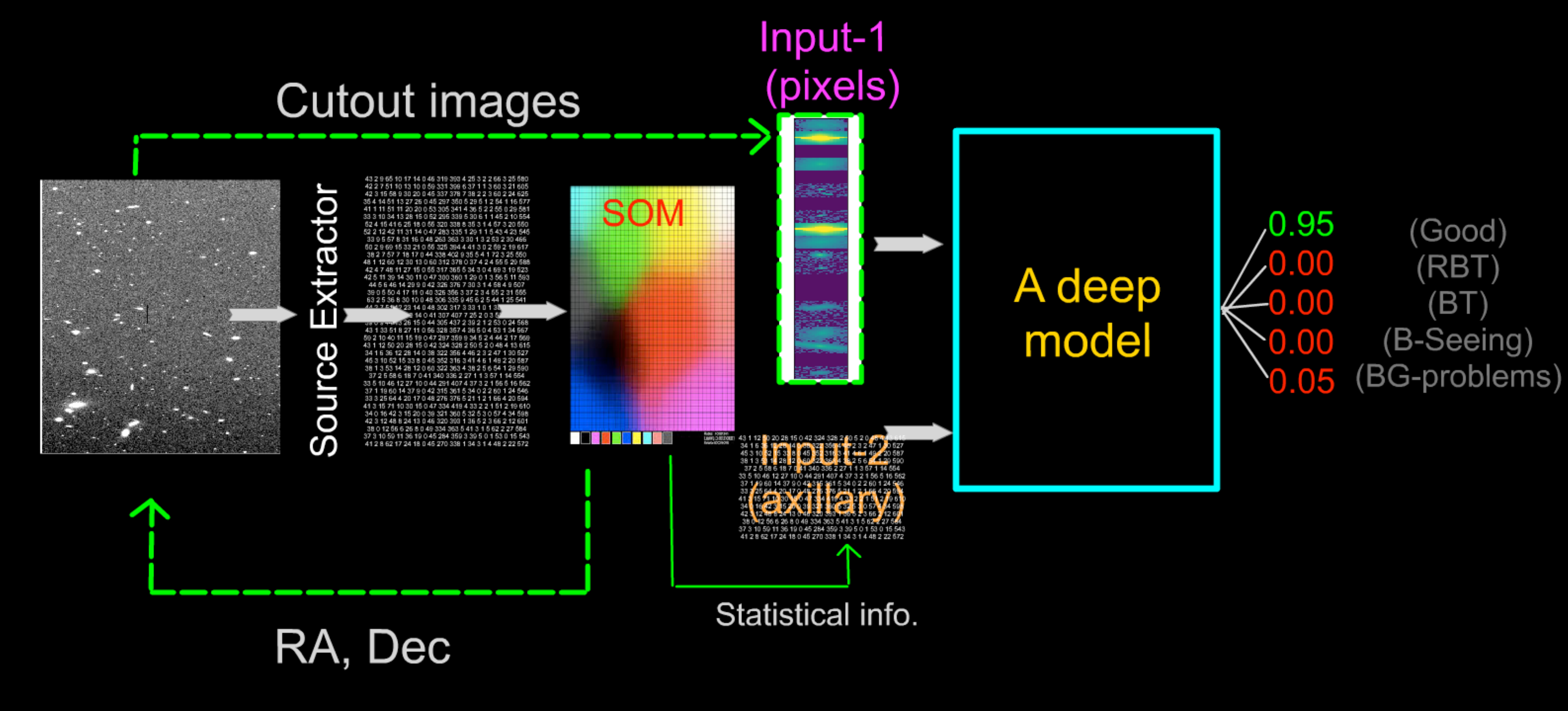}
\vspace{.5cm}
\caption{The plot is the combined model used in T20: “the selected parameter of detected sources from an image (the left image) can be extracted by SE, and the table from SE is then fed to a trained SOM. The SOM model can cluster the table into 20 clusters, and then we can pick up one object from each cluster. After that, a representative image (i.e., 20 cut-out objects using RA and Dec provided by SOM), using the main image, can be constructed. The representative image (i.e., pixel information) is the primary input (Input-1). Besides, we can obtain statistical information from SOM (the number of similar objects in different clusters). That is more information we will provide to the deep model (Input-2). The five classes are the output of the last model”.}
\label{fig:combined}
\end{figure*}

Figure \ref{fig:combined} shows the combined model, which is used in T20. First, they use the Source Extractor (SE) package \citep{bertin96}  to extract useful information concerning the 'sources' detected in the image. This information contains five parameters, such as the size of sources and the elongation (the left side of Figure~\ref{fig:combined}).  A trained SOM clusters the parameters. The SOM model places the sources into 20 clusters, and then one representative source from each cluster is selected. The 20 representative small images (pixel information) are shown as Input-1 in the Figure. There are available statistical data from SOM that can be used by the combined model (i.e., the number of similar objects in different clusters).  This information is denoted as Input-2 in the Figure. The deep model's target is the five different probabilities described above. The combined models, with two sets of Input-1 and Input-2, increase the performance significantly.

During the pre-processing procedure in T20, five SE parameters are needed to train the SOM in their pipeline.  In this paper, however, to train the first part of the pipeline (see Sec. \ref{sec:pipeline}), we only need to find the coordinates of a detected source in a CCD. A source is defined as five connected pixels (2$\sigma$ above the background). Therefore, no SE parameters, such as size or ellipticity of detected sources, are required here. Accordingly, any signal or source-detector, which can provide the detected sources' coordinates, can be used in the unsupervised method presented in this paper. These sources will then be used, by a deep model, to create a `comprehensive' prototype of the sources (see Figure \ref{DESOMmap} and Sec. \ref{sec:DESOM-1} for a description of how to create the prototypes).

\section{The method}
\label{sec:method}

In this section, we present our method. Since the final model consists of different ML bases, we will first describe them in a more detailed manner.

\subsection{Self-Organizing Maps}
SOM algorithms provide clustering and efficient data visualizations. A SOM is a feed-forward neural network with no hidden layers. There are therefore two layers, the input layer and the output map, which is typically a two-dimensional lattice with a defined number of units. The number of units is a hyperparameter. Proper selection of the number of units helps achieve higher accuracy and better performance. Each unit on the output map is a neuron represented by a prototype vector. Adjacent units on the map are related to each other through a neighborhood function. When an input data point is presented, the output units compete against each other so that the output unit whose prototype vector is closest to the input data wins. Eventually, the units that are close to each other on the output map, represent data points that are similar to each other in the input space.

The iterative training process starts with a random initialization where random values are assigned to prototype vectors of units. During each training step an input data point, from the training set, is selected. The distance between this data point and all prototype vectors is calculated. Euclidean distance is widely used in SOM.  The closest prototype, also known as the best matching unit, is selected as the winner. Next, the prototype vector of the best matching unit as well as all its adjacent units are updated as:

\begin{equation}
{w}_j(t+1)= {w}_j(t)+\alpha(t)\Lambda _i(r)[x(t)-{w}_j(t)]
\end{equation}

Where $w$, $x$ and $t$ are unit prototype vector, input data point, and time (i.e., the current training step), respectively. Subscript $j$ represents the $j^{th}$ prototype vector to be updated.   
$\Lambda_i(r)$ is the neighborhood function of the best matching unit, which is a function of its distance from its neighboring units. $\alpha(t)$ is the learning rate. Both the neighborhood function and the learning rate are a function of time, since they are updated and optimized as the algorithm continues. They both decrease monotonically with time. Usually, a Gaussian neighborhood function is used. It is an inverse exponential function of the distance of the best matching unit from all other units. The optimization goal is to jointly minimize the neighborhood kernel distance as well as the distance between input data points and the prototype vectors. The concept of neighborhood kernel is what differentiates a SOM algorithm from Kmeans. If the neighborhood kernel is set to 1 for the best matching unit and zero for all else, the algorithm will perform similarly to a Kmeans method. 

The SOM algorithm, unlike a Kmeans method, can efficiently consume large amounts of input data. However, the training time significantly increases as the map size increases. Several approaches are suggested to overcome this issue \citep[e.g., see ][]{Ivezic04,sompy18}.

\subsection{Autoencoders}
Autoencoders are a type of neural network that reconstructs the input data set in the output layer. They often have a symmetrical structure. In some types of autoencoder, the bottleneck contains the most compressed data representing the input data set. The learning process that compresses the input data set is called the encoder, and the process to reconstruct the input data set from the bottleneck layer is called a decoder. Since the input data is produced at the output layer, the number of neurons in the input and output layers are equal. Hidden layers might have a different number of neurons depending on the problem and the input data set. There are several different types of autoencoder, such as sparse or deep autoencoders \citep[e.g.,][]{Zhang18}. Autoencoders can be grouped based on various criteria such as their structure type, learning algorithm, or type of loss function. However, the most common way to group them is their structure, i.e., the number of neurons or the number of layers. Autoencoders can be either undercomplete or overcomplete. An undercomplete autoencoder has fewer neurons in its bottleneck layer, while in an overcomplete autoencoder, the number of neurons in the middle layer could be equal to or more than that of the input layer. Proper conditions and restrictions need to be applied to overcomplete autoencoders to prevent them from copying data from the input layer to the output layer.
Additionally, autoencoders can be either deep or shallow, depending on how many hidden layers they have. Various types of autoencoders are used to solve different problems. Here we focus on undercomplete deep autoencoders.

At a high level, an autoencoder function could be considered as follows:

\begin{equation}
x' = g(f(x))
\end{equation}

Where $x$ and $x'$ are the input data and the reconstructed data, respectively. $f$ is the encoding function and $g$ is the decoding function. The goal is to make $x'$ as close to $x$ as possible. 
Various loss functions can be used as objective functions, such as least square errors or root mean squared errors.  To optimize the weights and bias factors of the neurons, different algorithms such as stochastic gradient descent (SGD) can be used. During this algorithm, at each step the gradient of the objective function is calculated with respect to each parameter. The goal is to move in the direction of the global minimum of the function by taking into account a learning parameter. The back-propagation technique is used to obtain the gradients at each layer. Various regularization methods are discussed in the literature to prevent overfitting and to prevent the vanishing gradient problem in autoencoders \citep{Ioffe15}.

If the input data is an image, convolutional autoencoders are the most efficient classification method. This class of autoencoders incorporates the structure of a convolutional neural network (CNN) \citep[e.g., see ][recent advances in CNN]{Gu15}. A CNN is a modified traditional neural network that is typically deeper and wider. While traditional neural networks directly consume pixel data points, a CNN first transforms the input pixels through data processing layers, then feeds them into a fully connected neural network. The processing layers act as a feature-learning procedure where various important features and patterns within the input data are detected and extracted. This learning process also reduces the input image size. Therefore, the data set sent to the fully connected neural network is denser and more compact, consisting of important input data features rather than all pixel values, contributing to a more efficient algorithm. The data processing layer usually consists of various filters. A filter is an $n\times n$ matrix that scans through the input image pixels and then converts each $n\times n$ pixel patch of the input image into a number based on certain calculations. For example, a filter known as “max pooling” selects the highest pixel value within each $n \times n$ pixel patch. This procedure allows proper pattern-detecting \citep[e.g., see][for a more detailed explanation about the max pooling function, regularization, and CNN]{Wu15,Teimoorinia20a}.

An important type of autoencoder is the variational autoencoder(VAE) \citep[e.g., see][for a comprehensive introduction]{Kingma19}. Rather than reconstructing the input image, as traditional autoencoders do, a VAE learns parameters of the probability density function representing the input data set. The continuous latent space of a VAE allows the algorithm to randomly sample from the latent space to generate variations of the input image. 

In summary, the bottleneck layer of an undercomplete autoencoder contains the compressed representation of the input data set. This low-dimensional data can be efficiently used in clustering algorithms that otherwise fail to process high-dimensional data sets. This is further explained in the next section.

\subsection{Deep SOM}
\begin{figure*}
\centering
\includegraphics[width=14.cm,height=6.5cm,angle=0]{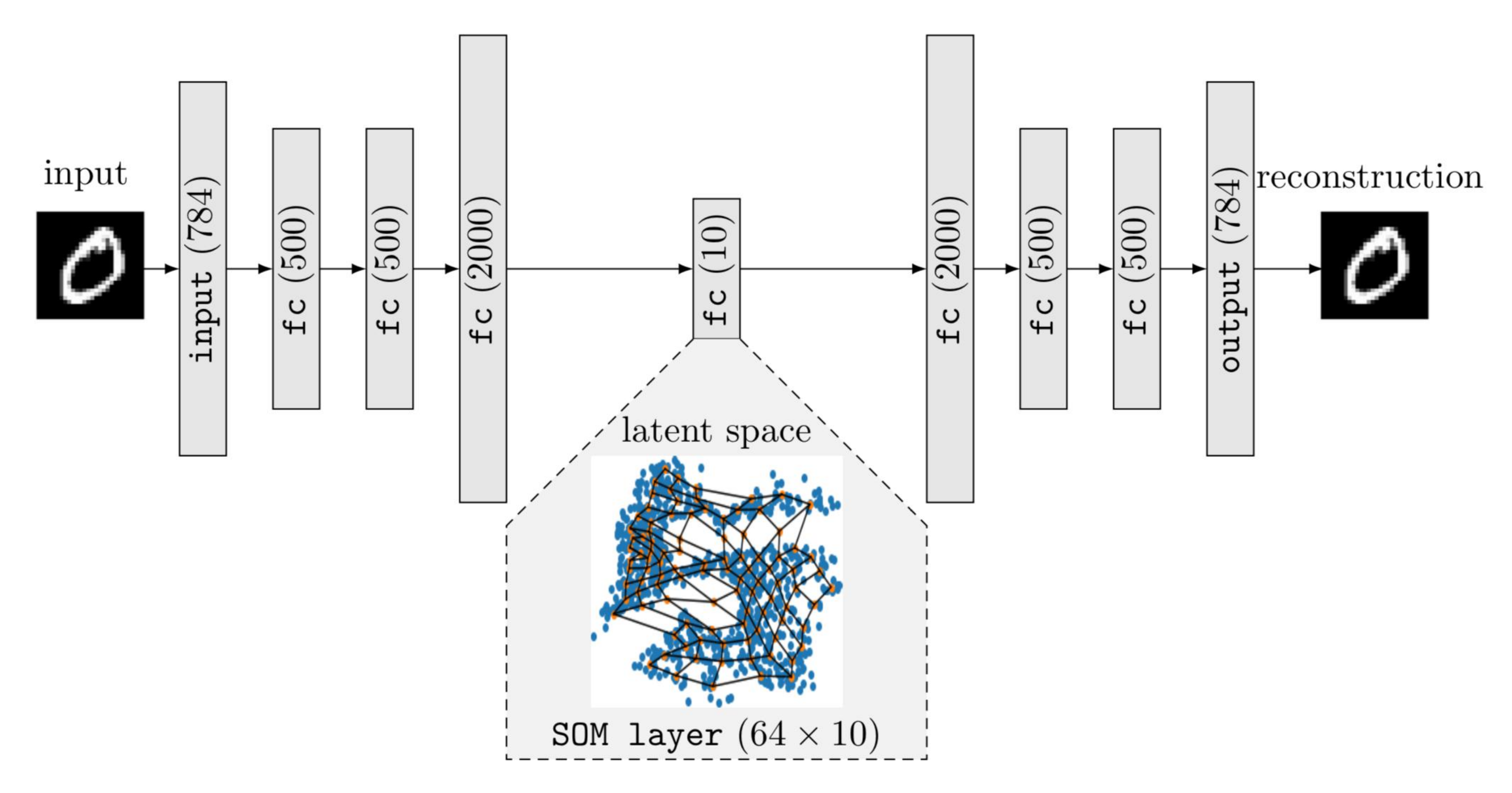}

\caption{Architecture of DESOM Model with Convolutional autoencoder [\citet{Forest2019}]}
\label{ConvDESOM}
\end{figure*}

A deep embedded SOM (DESOM) jointly trains an autoencoder and a SOM algorithm. The dense layer of the autoencoder provides efficient, low-dimensional input data for the SOM algorithm resulting in more efficient SOM performance. Furthermore, this iterative joint training improves the overall classification performance while reducing the training time. 
The architecture of this method, shown in Figure \ref{ConvDESOM} \citep[as an example from][]{Forest2019} embeds the SOM algorithm within the bottleneck of the autoencoder, the latent space. 

The SOM consists of $K$ prototype vectors described by $\left\{\mathbf{m}_k\right\}_{1\leq k \leq K}$. A Gaussian function can be used as the neighborhood function. The loss function of the joint method (see Figure \ref{gradients}) is defined below:

\begin{equation}
L (\mathbf{W_e}, \mathbf{W_d}, \mathbf{m_1},... \mathbf{m_K}, \chi) \\
= L_r(\mathbf{W_e},\mathbf{W_d}) + \gamma L_{SOM}(\mathbf{W_e}, \mathbf{m_1},...,\mathbf{m_K}, \chi)
\end{equation}

where subscripts $e$ and $d$ represent the encoder and decoder, respectively. $L_r$ is the least square loss of the reconstruction, the decoder. It is a function of decoder and encoder weight parameters. $L_{SOM}$ is the SOM loss—a function of encoder weight parameters, prototype vectors, and $\chi$, which is called the assignment function. It selects the best matching unit. $\gamma$ is a parameter that defines the trade-off between the SOM loss and the decoder loss.

\subsection{DESOM-1}
\label{sec:DESOM-1}

The labeling pipeline proposed in this paper consists of two sections: DESOM-1 and DESOM-2. DESOM-1 is composed of a CNN autoencoder that is jointly trained with a SOM (i.e., Figure \ref{ConvDESOM}). The CNN autoencoder has a symmetrical structure: an input layer, three hidden layers within the encoder section, a dense middle latent layer, three hidden layers within the decoder section, and the output layer. A sequence of $3\times3$ filters followed by a $2\times2$ max pooling method is applied within the encoder. Next, data is flattened and then fed to a fully connected layer, connected to a dense middle layer \citep[e.g., see][for more details about pooling and flattened layers]{Teimoorinia20a}. Subsequently, the decoder layer reconstructs the input images by applying a two-dimensional up-sampling method within the two-dimensional convolution layers. A sigmoid activation function \citep[e.g., see][]{Nwankpa18} is used in the last layer. The autoencoder's dense layer provides a vector of size 160, which is the length of the prototype vectors in the SOM. This data is the input to the SOM, whose output is a $25\times25$ grid.

First, the autoencoder structure is defined and constructed. Then the prototype vectors of the SOM are initialized. After that joint optimization starts as described in \cite{Forest2019}. The weights of the encoder and decoder, $\mathbf{W_e}$ and $\mathbf{W_d}$, and SOM prototype vectors, $\mathbf{m_k}$, are optimized through a joint training procedure using a minibatch SGD algorithm.

A Gaussian neighborhood function is used in the SOM as follows: 

\begin{equation}
K^{T}(d) = e^{\frac{-d^2}{T2}}
\end{equation}

where $d$ is the topological distance between prototype vectors and $T$ is a temperature parameter that controls the radius of the neighborhood. At each training step, $T$ decays exponentially with respect to its user-defined minimum and maximum range. Care must be taken in selecting the appropriate temperature range. If the minimum temperature is too low, the SOM will converge to a Kmeans algorithm that may suffer from the curse of dimensionality.

In our model, the temperature parameter minimum is 0.5, and its maximum value is 10. 100,000 iterations are used for training, and the trade off between the SOM loss and the decoder loss, $\gamma$ is $0.0075$.

The training procedure can be summarized as follows as described in \citep{Forest2019}:
\newline

Initialize AutoEncoder (Glorot uniform) 

Initialize SOM parameters (random samples)

for iter = 1, ..., iterations do 

\hspace {\parindent} \begin{math}T \leftarrow T_{max} (\frac {T_{min}}{T_{max}}) ^ {\frac{iter} {iterations}}\end{math}

\hspace {\parindent} Load next training minibatch 

\hspace {\parindent} Predict SOM pairwise distances on minibatch 

\hspace{\parindent} \hspace {\parindent} and compute weights \begin{math}w_{i,k}\end{math}

\hspace {\parindent} Train DESOM on minibatch
   
End

Output AutoEncoder weights $\mathbf{W_e}$, $\mathbf{W_d}$ 

\hspace {\parindent} and SOM prototype vectors $\mathbf{m_k}$
\newline

\begin{figure*}
\centering
\includegraphics[width=12.cm,height=5cm,angle=0]{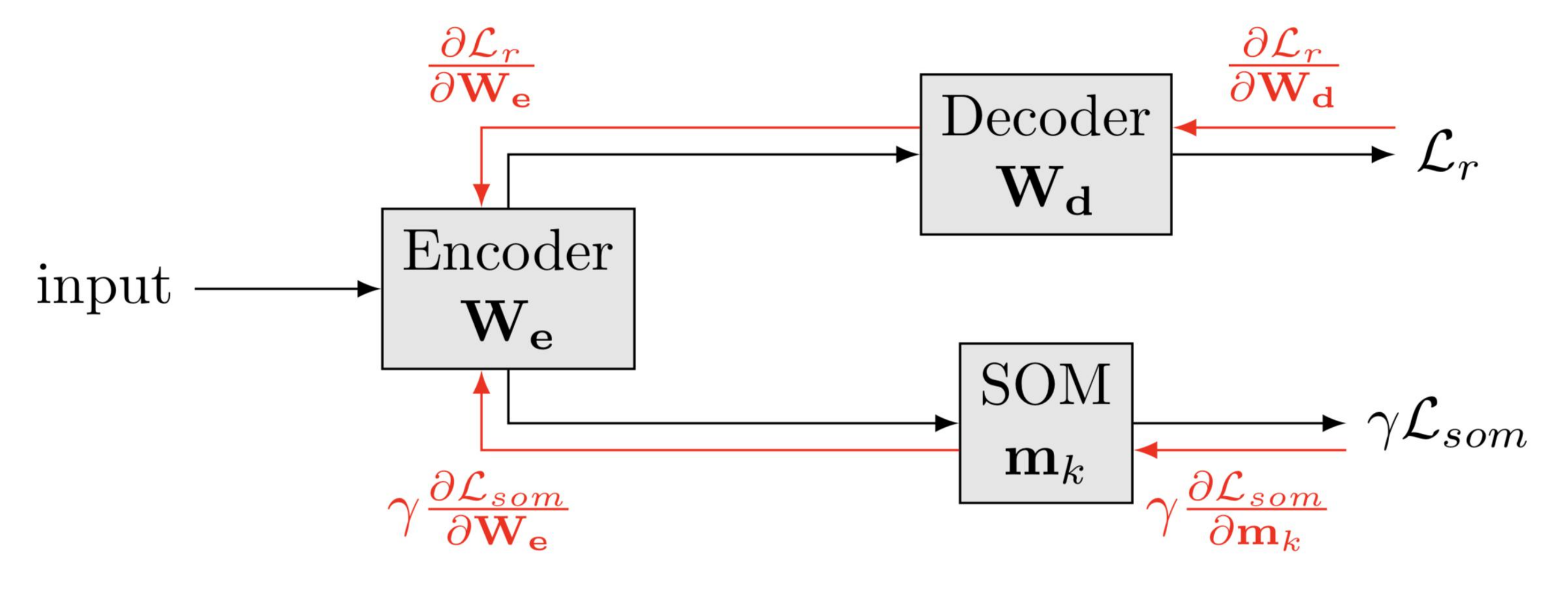}
\caption{Path of Gradients in the DESOM Model [\citet{Forest2019}]}
\label{gradients}
\end{figure*}

The paths of the gradients of the loss function are illustrated in Figure \ref{gradients} \citep[from][]{Forest2019}. 

Minibatch SGD computes the gradient of the loss function with respect to the network parameters for each minibatch of $n_b$ training samples.

\begin{figure*}
\centering
\includegraphics[width=18cm,height=18cm,angle=0]{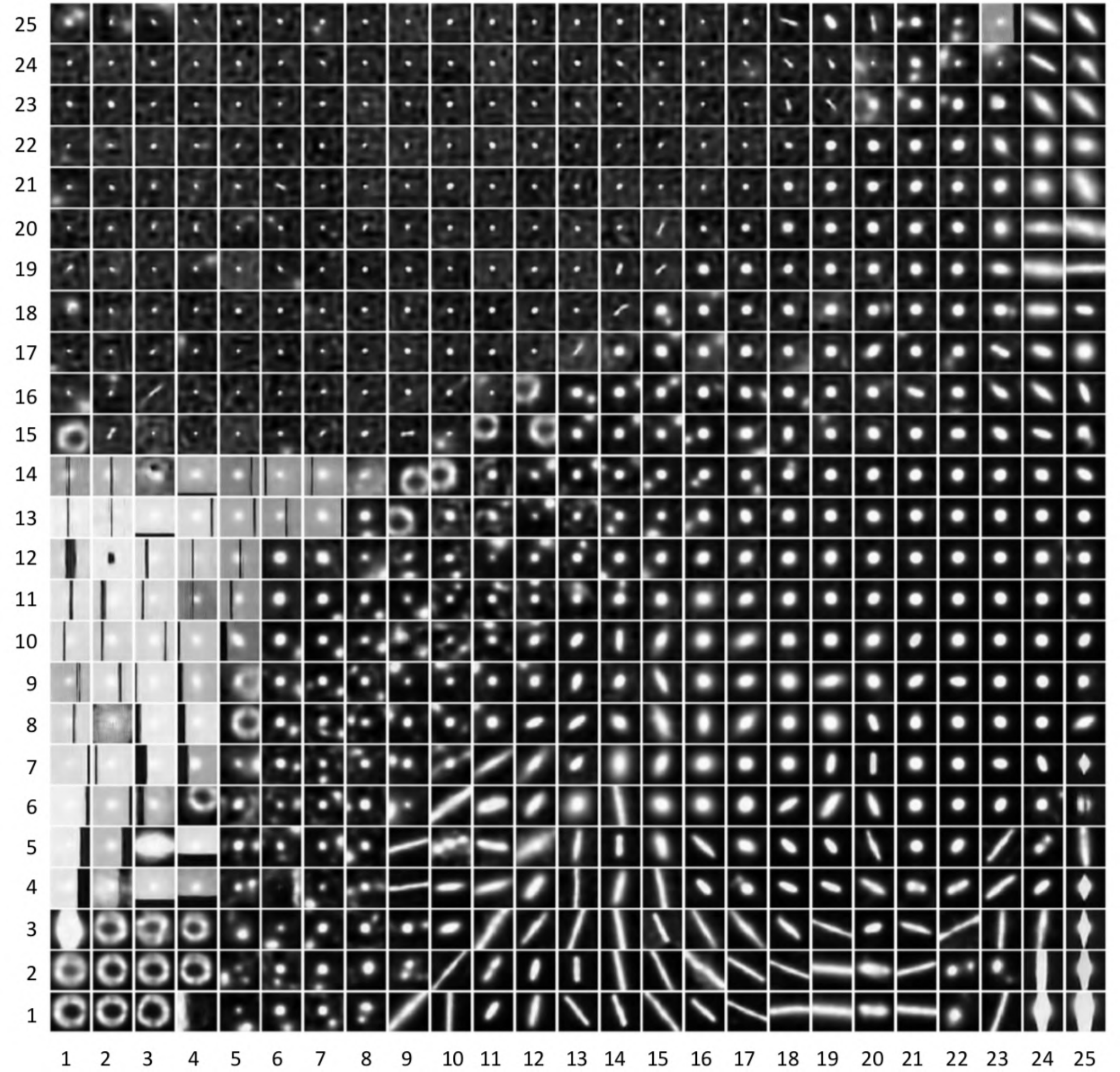}

\caption{Output map of DESOM-1 algorithm}
\label{DESOMmap}
\end{figure*}

A significant output of the model (DESOM-1) implementation is a visual representation of the map of prototype vectors. The map generated by our DESOM-1 is shown in Figure \ref{DESOMmap}, which is a good representation of objects in the training set. This presentation is extremely useful to visually assess and identify the types of objects present in the dataset.  It can be seen that the algorithm identifies various shapes, rotations, sizes, and imaging conditions that are present in the entire training dataset. It should be noted that they are not real astronomical objects such as galaxies or stars. In fact, they have been created during the training step. In the following section, we will describe the DESOM-1 training step and the useful outputs.

\subsection{Training DESOM-1}

DESOM-1 can be used to identify image characteristics within the training set. Depending on the types of objects present and the image's quality, the training set creates a certain pattern on the SOM map. In this paper, we train a SOM ($25\times25=625$ nodes) using $1,000,000$ cutout images randomly selected from the $\sim70000$ MegaCam exposures described in Sec. \ref{sec:data}. Each cutout image has a size of $32\times32$ pixels. We use Keras \citep{keras}, a deep learning API running on top of the Tensorflow platform \citep[][]{tensorflow2015-whitepaper}. Keras has a SOM layer that makes it possible to use GPU facilities when we train DESOM-1. Using a GPU, we train DESOM-1 with one million $32\times32$ images in less than 3 hours. Generally, the choice of a training set size can depend on the problem's complexity under study. Megacam images are complicated with a large variety of objects. Training DESOM-1 could be done with CPU processors, using fewer images. For a map of size $25\times25$, using a sampler sample such as 200K cutouts images, does not show any `visually' significant difference in the maps. However, to obtain a `complete' set of representative prototypes and also to avoid possible issues such as overfitting, we take a considerably sizable input to train DESOM-1. On the other hand, since the input to DESOM-2 is not images, training DESOM-2 is not very time-consuming with CPU processes (typically a few hours,  depending on the processors' speed).

The trained DESOM-1 ($25\times25$) shows 625 prototypes in Figure \ref{DESOMmap}. These prototypes are representations of potential cut out objects in Megacam exposures. For example, different RBT objects can be seen in different sizes, elongations, and orientations, e.g., nodes (9, 1) and (15, 1). As another example, the round objects with a hole in them, e.g., node (3, 2), represent bad focus characteristics. We chose a SOM with the size of ($25\times25$). If we had chosen a small map, it could have hidden potential scenarios and information on the map. Conversely, choosing an extensive map could have been costly \citep[e.g., see][]{Ivezic04, Teimoorinia20a}. We increased the map size during the training steps to see a reasonable sample in terms of object orientation, size, and shape.

The structure of CNNs make predictions invariant to small translations such as rotation in some pattern recognition tasks like classifications. However, Figure \ref{DESOMmap} shows different objects' rotations, indicating that DESOM-1 is not invariant under different rotations. In this respect, it should be noted that we aim to reconstruct an image of an object in the autoencoder part of the system. Therefore, the latent data should have the original image's encoded information, including the object's orientation. This encoded information is then used by a SOM that is not an invariant rotation system \citep[e.g., see][]{Polsterer15}.

If we were to use the trained DESOM-1, then any new single cutout image of size $32\times32$ (as a single test input to the trained model) should match one of these 625 prototypes (nodes). In other words, any new cutout image would have to “land” somewhere on the DESOM-1 map. Suppose we collect N cutout images from a selected MegaCam exposure (as the input to trained DESOM-1). Then, the N images will be distributed on the map somehow. This distribution is an example of a “density map” related to the exposure. Each cell within the map might contain multiple objects suggesting that they are similar. Here N is the detectable source in the exposure, and any source- or signal-detector software package may do this step during pre-processing. In other words, we just need to find a signal in an exposure and cut the associated source out of it. The important point is that the character of a density map depends on the nature of the selected exposure. 

\begin{figure*}
\centering
\includegraphics[width=16cm,height=16cm,angle=0]{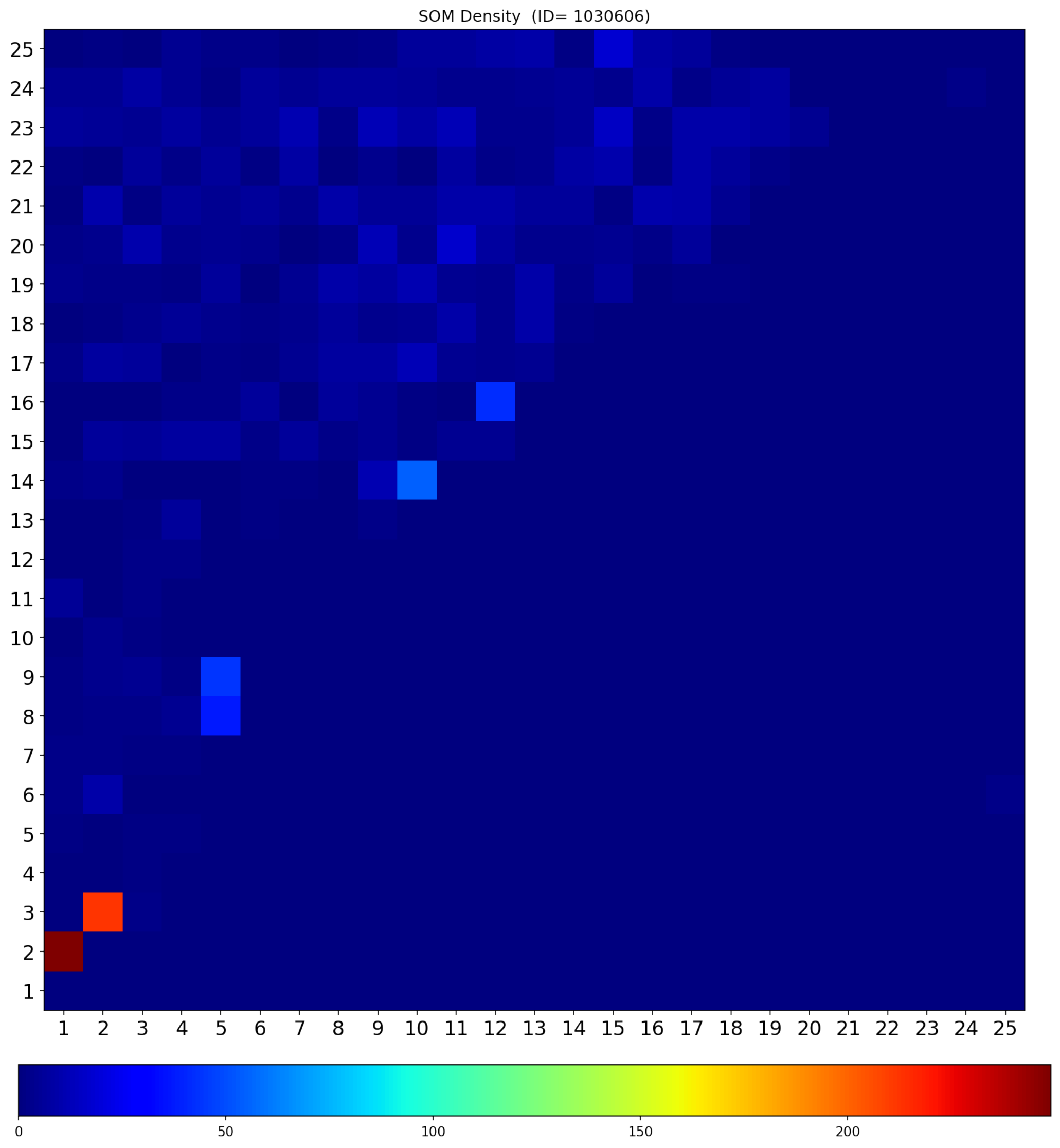}
\caption{An example of a density map related to a bad focus exposure. Nodes (1, 2) and (2, 3) are comparatively crowded. The same nodes in DESOM-1 (Figure \ref{DESOMmap}) show the nature of the associated objects.}
\label{fig:SOM-Dnsity}
\end{figure*}

For example, an RBT image's density map should be very different from a “Good” one.  The density map from the RBT exposure should mostly be populated by the prototypes that show elongated shapes, because objects in the RBT should generally have an elongated characteristic, e.g., node (14, 1) in Figure \ref{DESOMmap}, whereas a Good image should show a more diverse population, and the cutout images (from the Good image) should be more uniformly distributed across the associated density map. As another example, a set of cutout images from a bad focus exposure should have a high density around node (2, 2) in the map, where round images with a hole inside are presented. Figure  \ref{fig:SOM-Dnsity}, as a detailed example, shows a density map of bad focus exposure.  As can be seen, nodes (1, 2) and (2, 3) are comparatively crowded. The same nodes in DESOM-1 (Figure \ref{DESOMmap}) show the objects' nature. So, we have a content-based recommendation system that can have different applications (see Sec. \ref{sec:discussion}).

\begin{figure}
\centering

\includegraphics[width=5.5cm,height=5.5cm,angle=0]{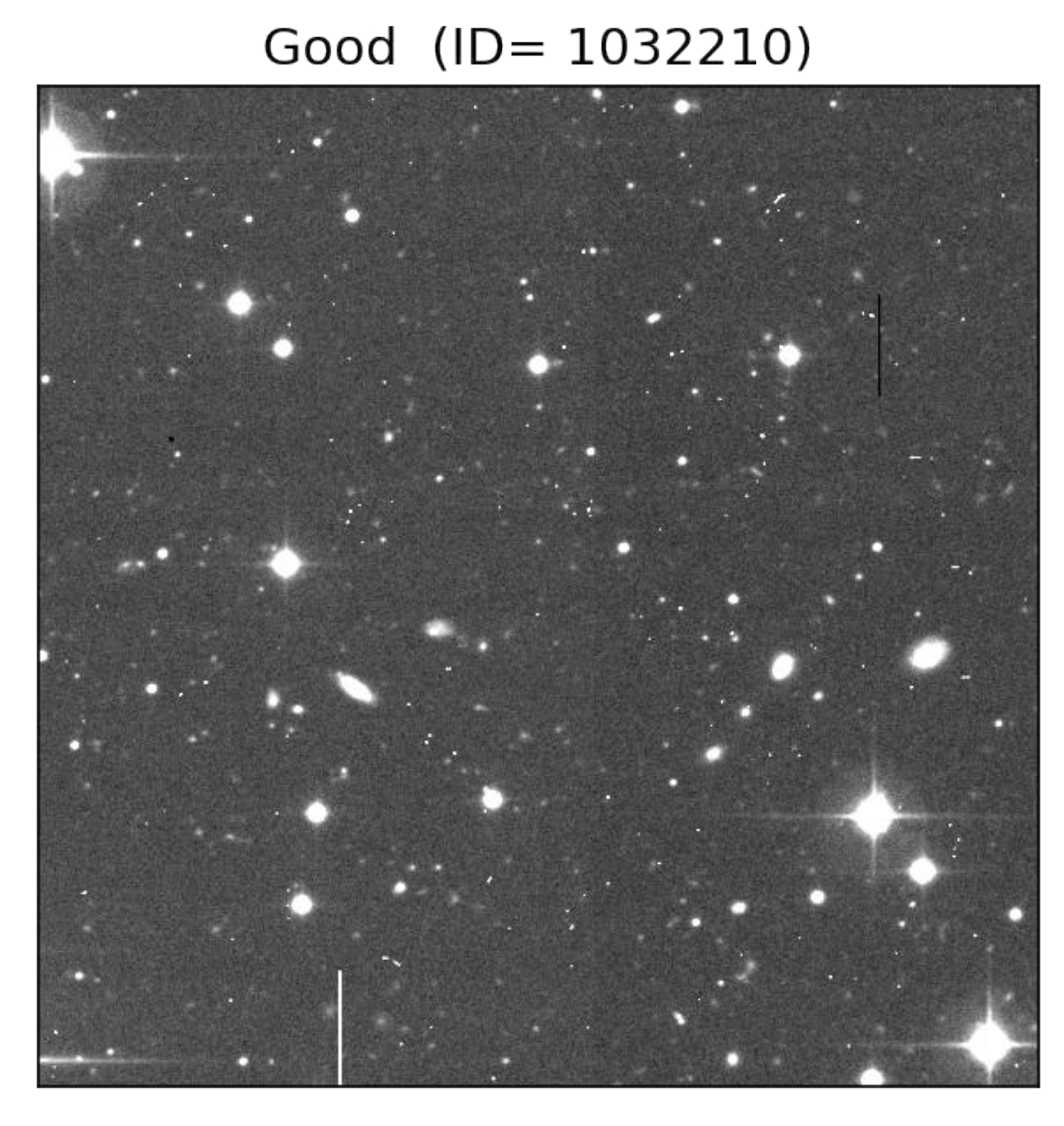}
\includegraphics[width=5.5cm,height=5.5cm,angle=0]{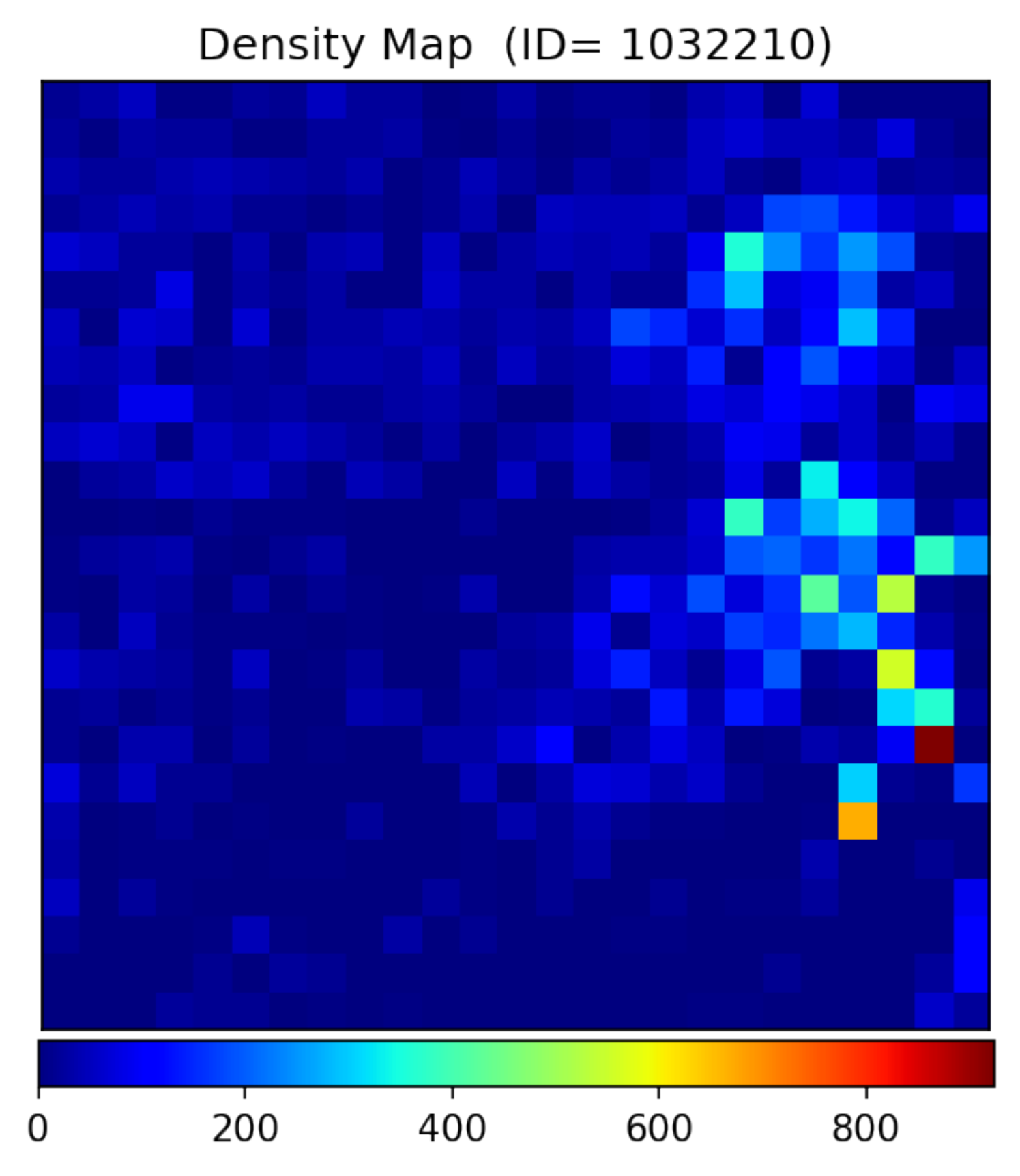}
\includegraphics[width=5.5cm,height=5.5cm,angle=0]{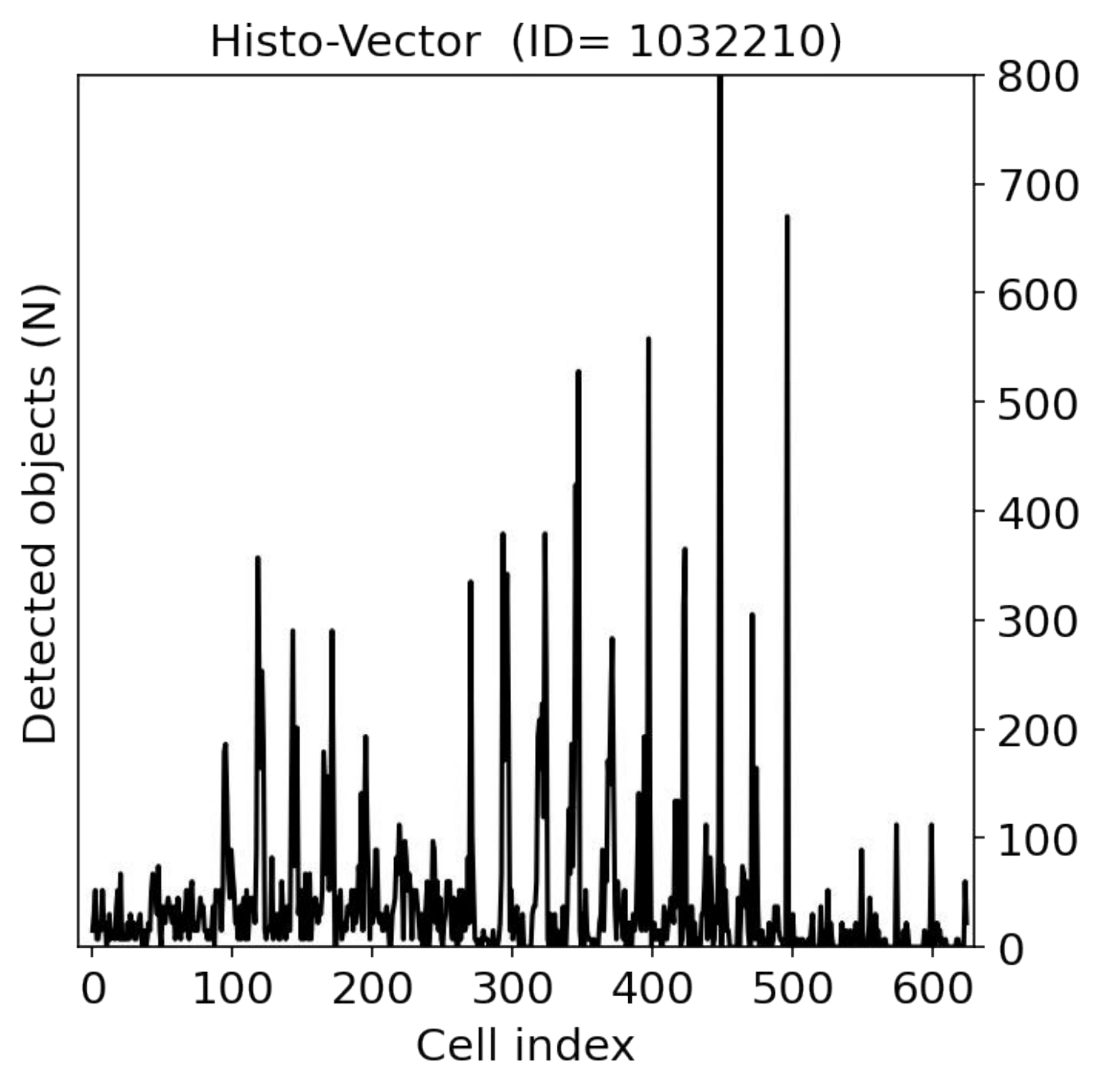}

\includegraphics[width=5.5cm,height=5.5cm,angle=0]{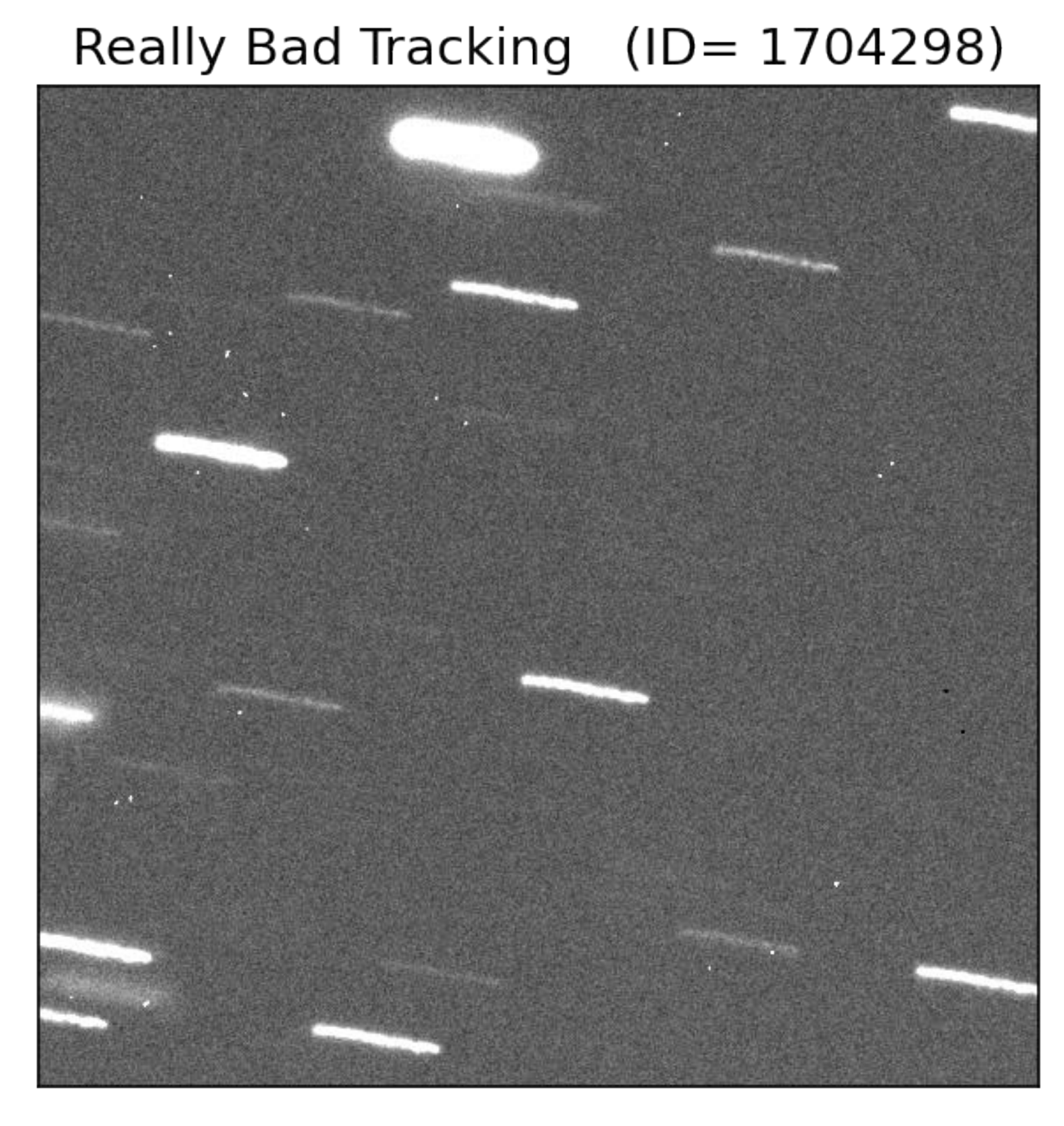}
\includegraphics[width=5.5cm,height=5.5cm,angle=0]{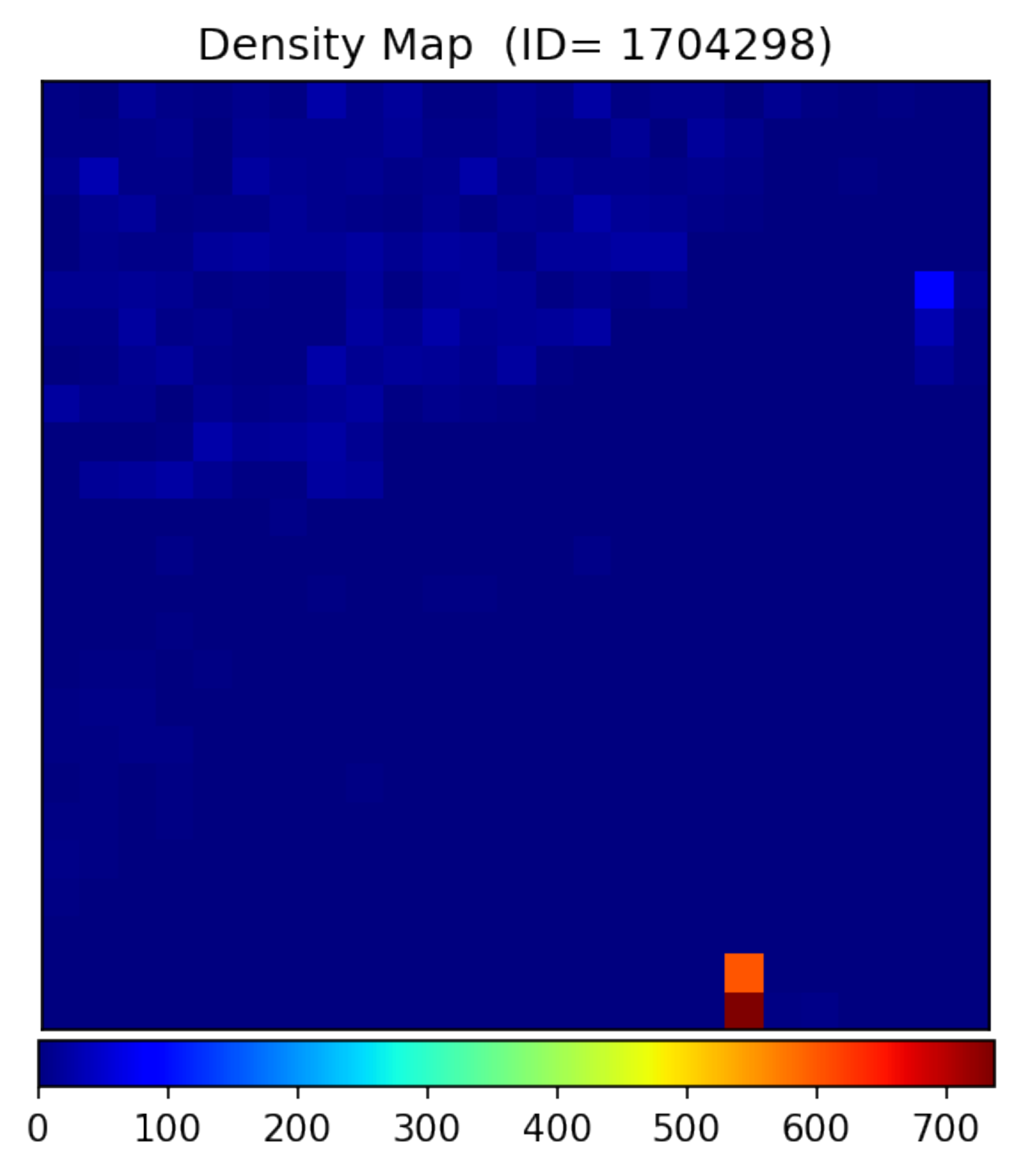}
\includegraphics[width=5.5cm,height=5.5cm,angle=0]{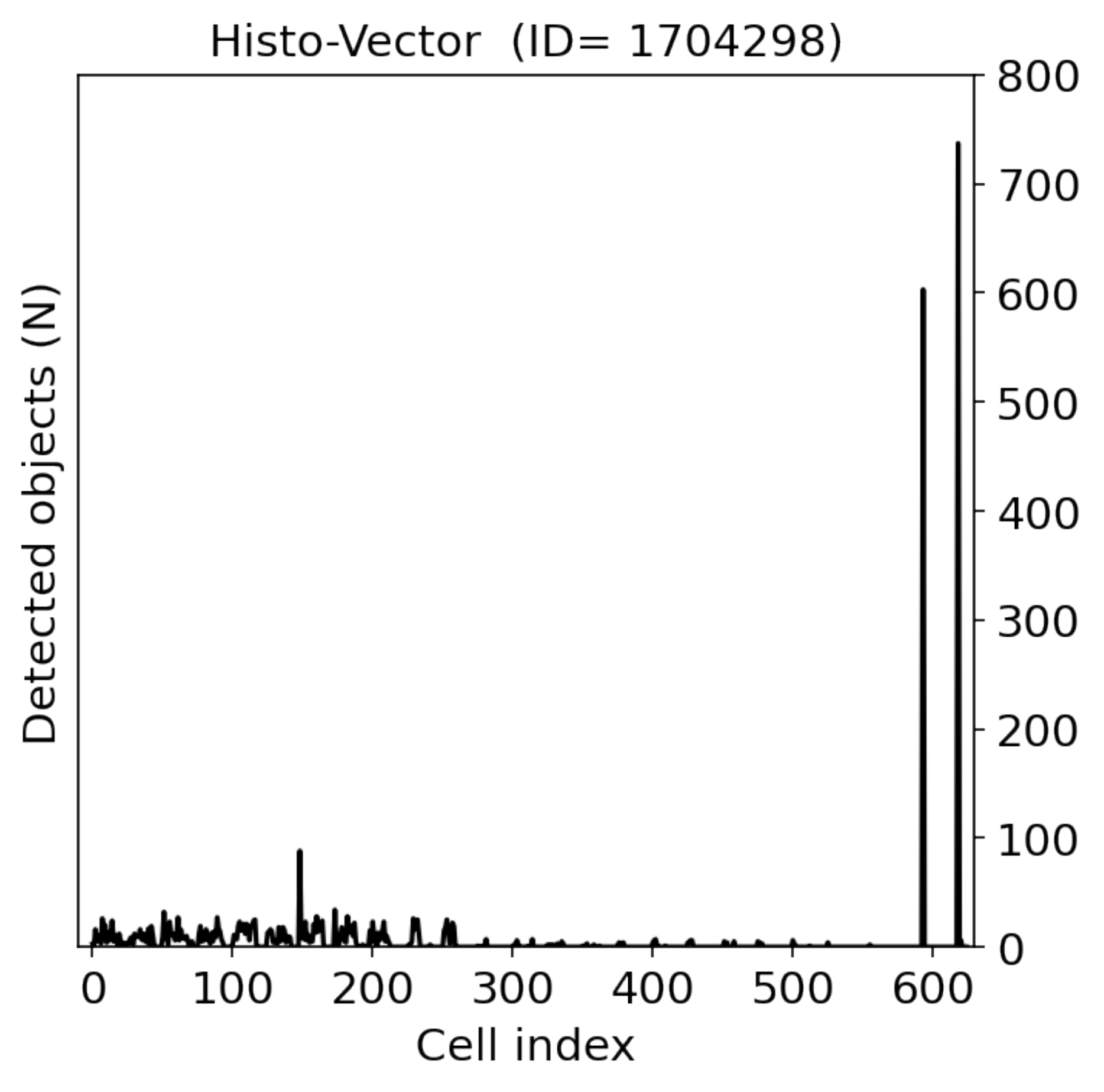}

\includegraphics[width=5.5cm,height=5.5cm,angle=0]{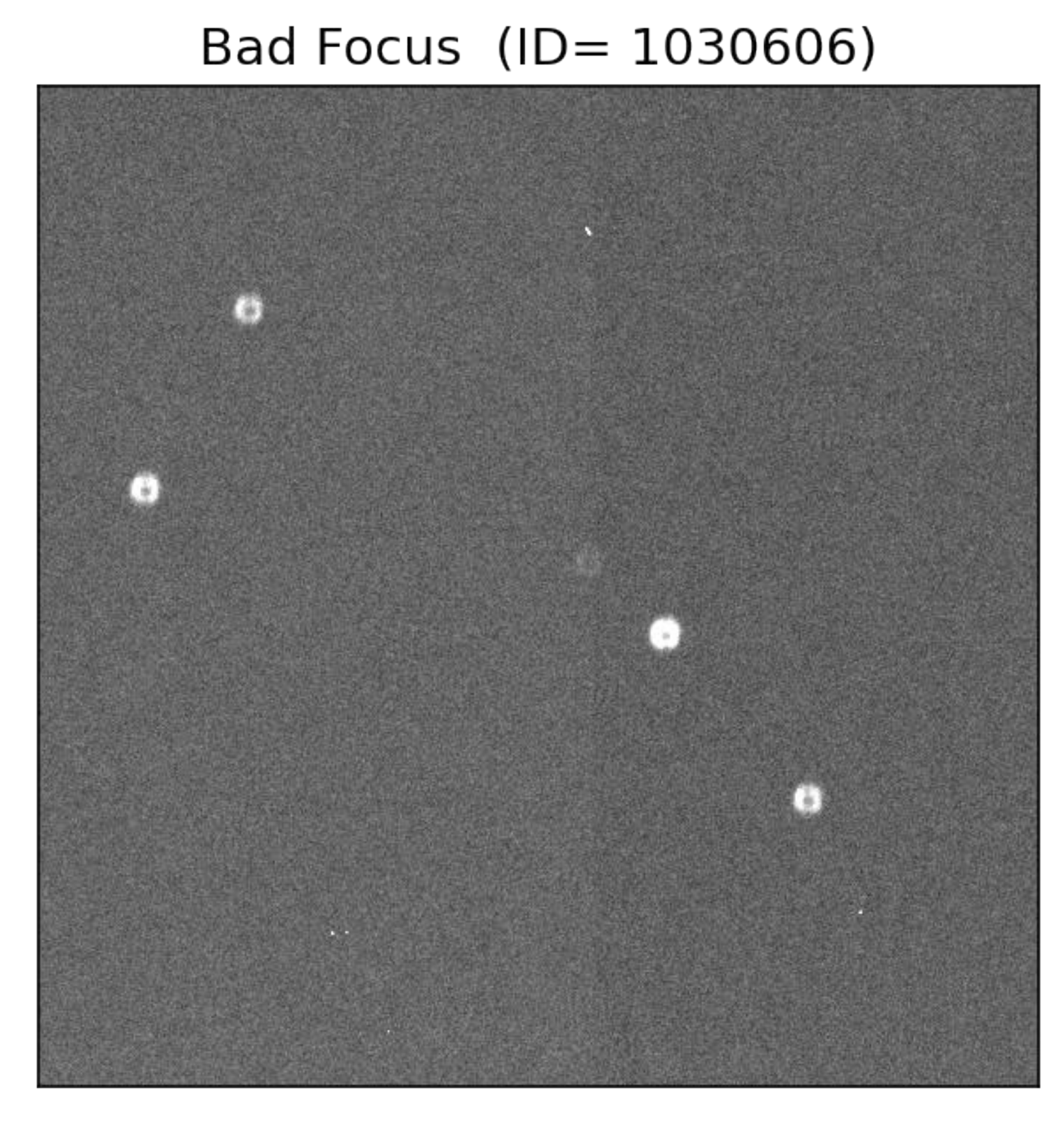}
\includegraphics[width=5.5cm,height=5.5cm,angle=0]{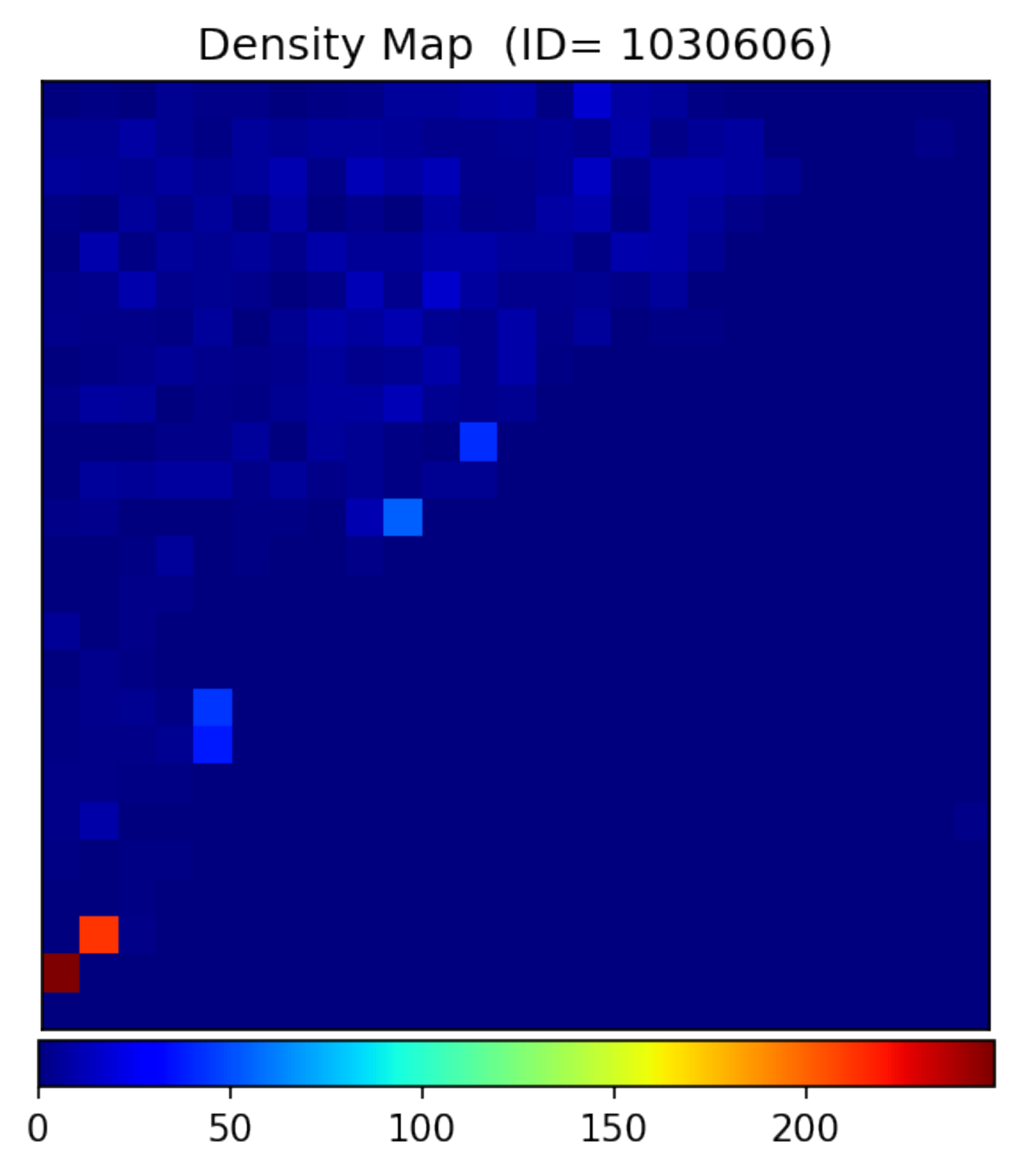}
\includegraphics[width=5.5cm,height=5.5cm,angle=0]{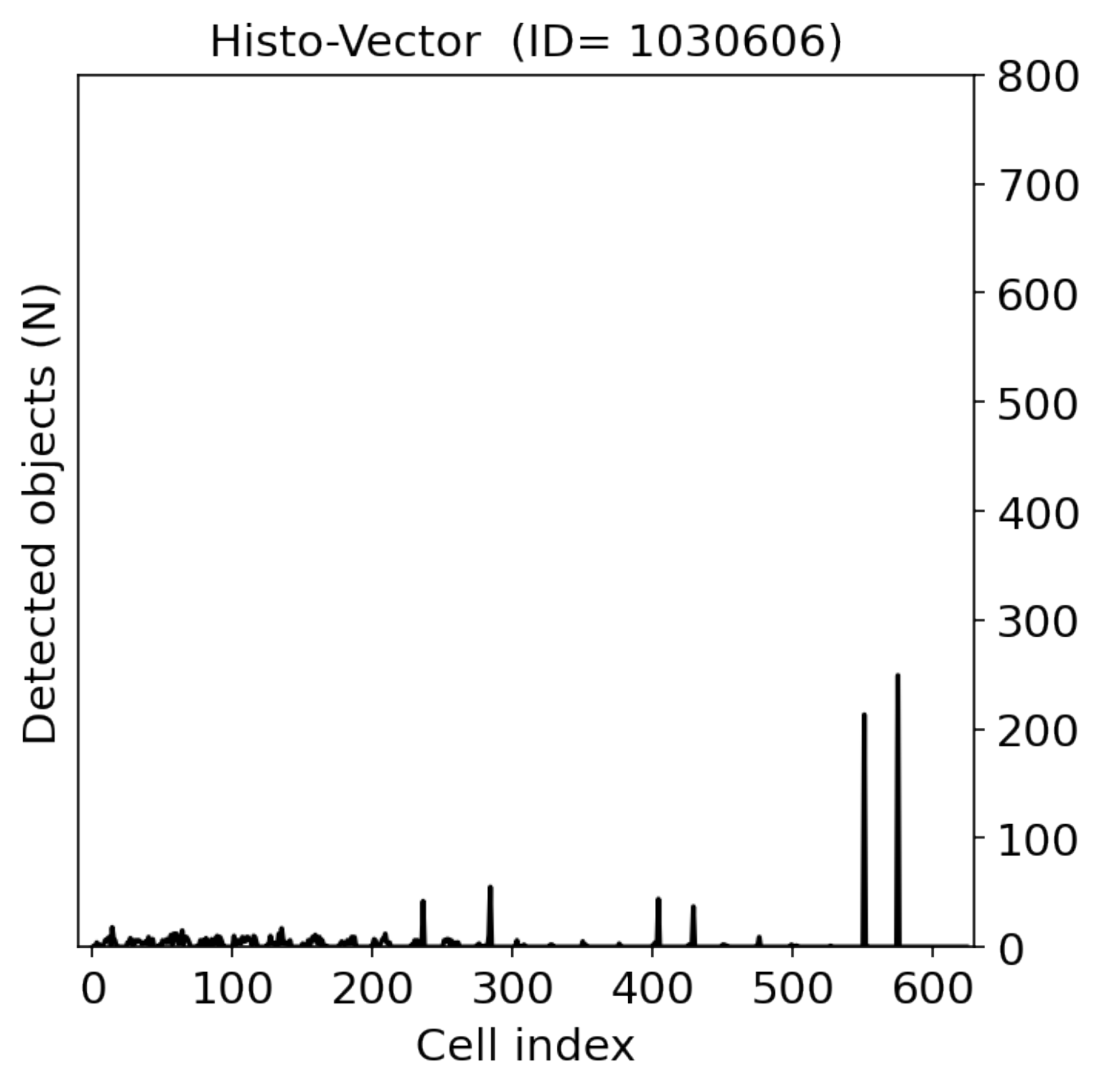}

\includegraphics[width=5.5cm,height=5.5cm,angle=0]{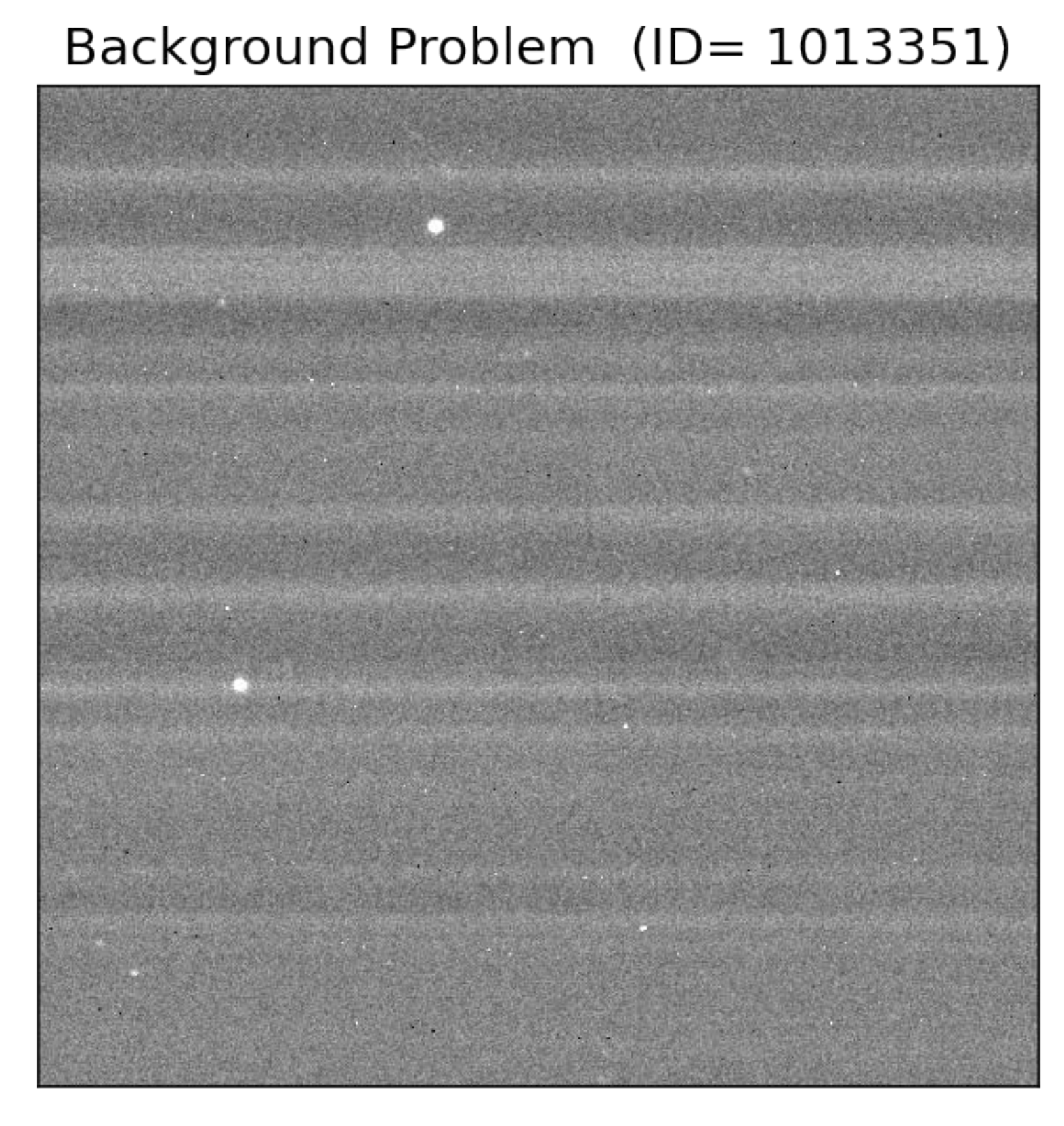}
\includegraphics[width=5.5cm,height=5.5cm,angle=0]{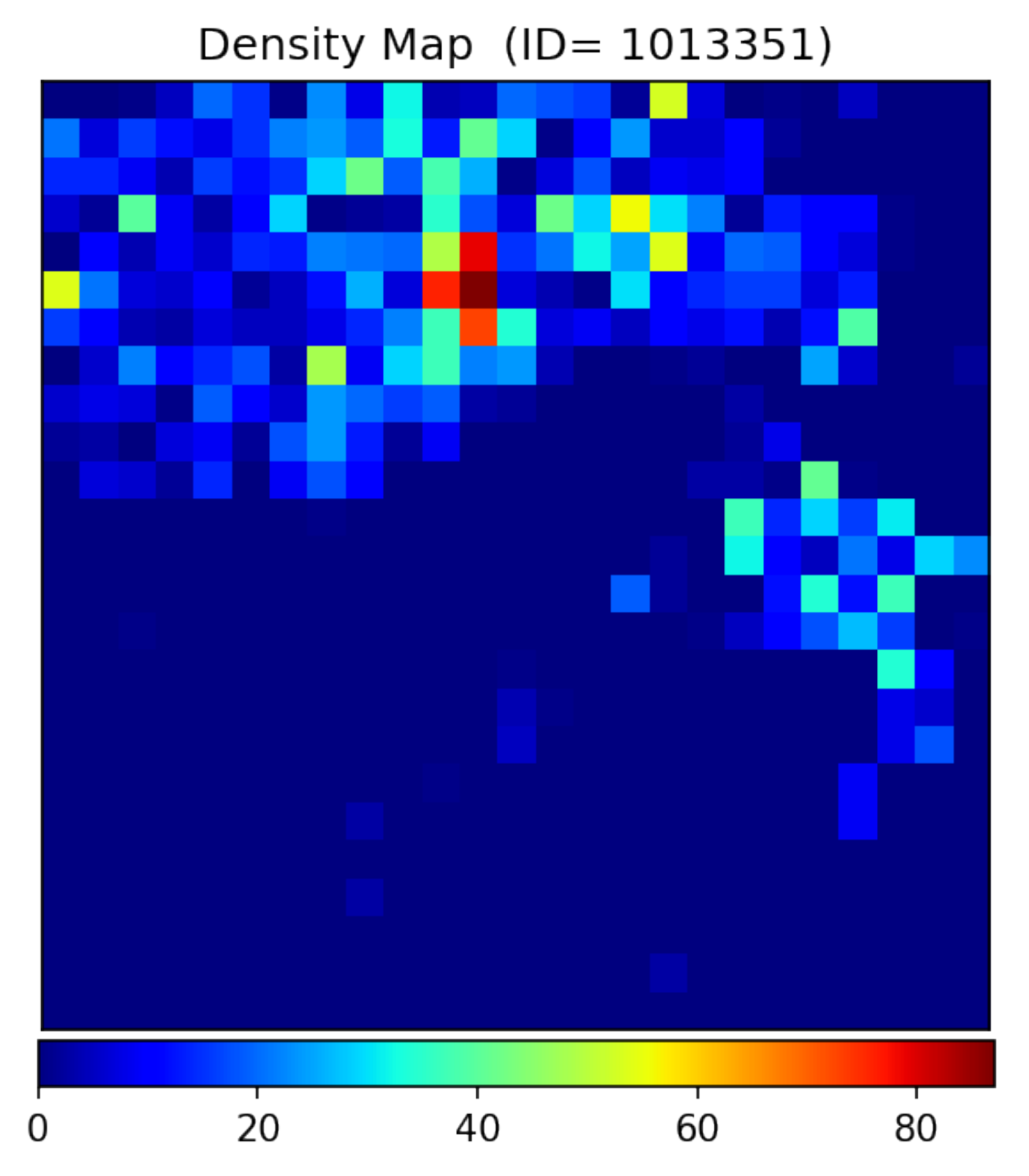}
\includegraphics[width=5.5cm,height=5.5cm,angle=0]{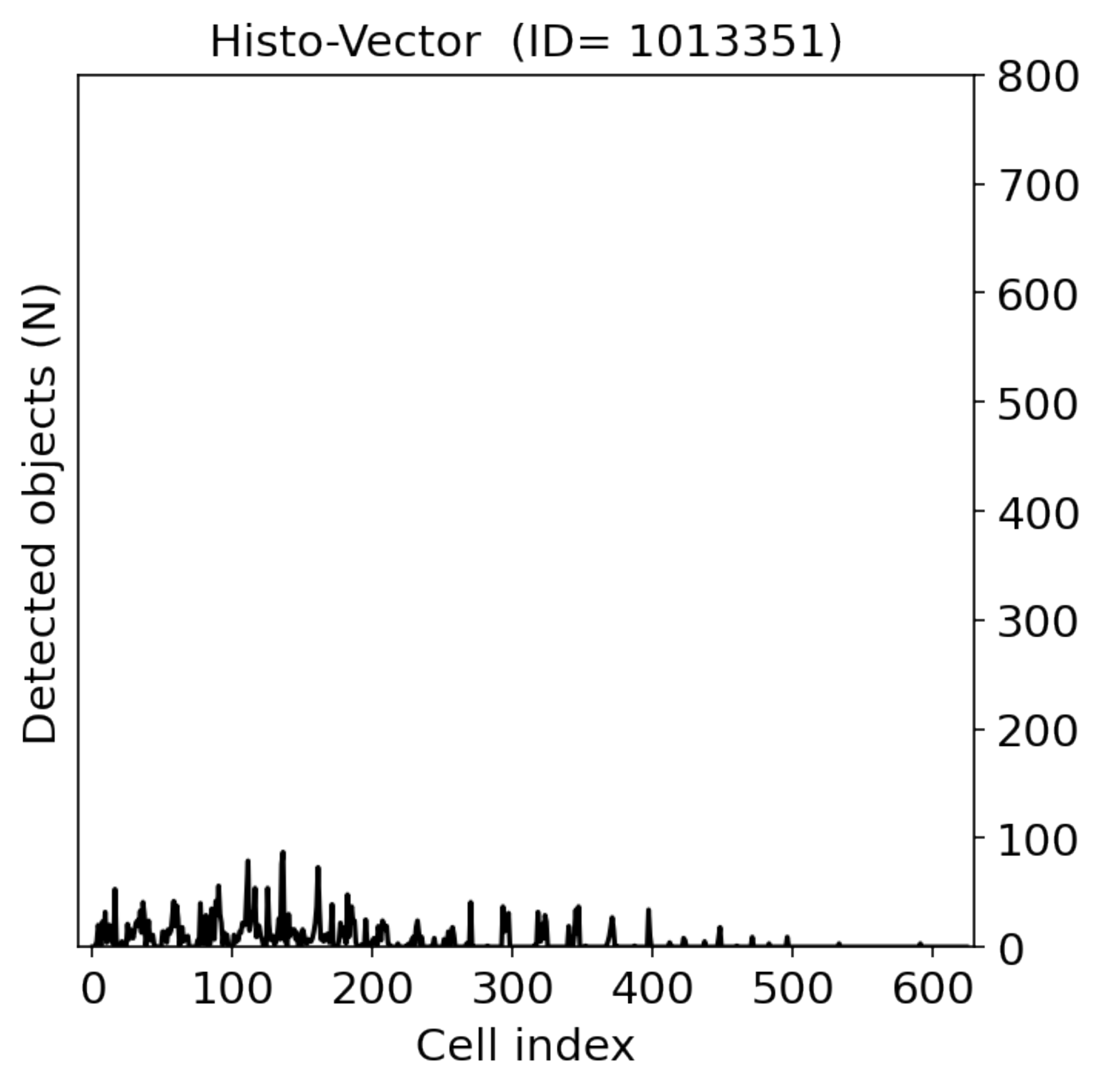}

\caption{From top to bottom: four exposures, each with different qualities and characteristics. For example, the top left shows a $500\times500$ sample from the much larger exposure 1032210. The top middle shows the associated density map, and the top right shows the associated histo-vector. As can be seen, the four IDs show different patterns (considering SOM densities or the histo-vectors). For better content-based matching, each SOM density here should be compared to Figure \ref{DESOMmap}.}
\label{fig:dense}
\end{figure}

As some more examples, in Figure \ref{fig:dense}, from top to bottom we show four different exposures with different qualities and characteristics. The left panel represents four images (size $500\times500$ pixels) of different exposures. The middle panel shows the associated density maps, i.e., the density of detected objects. We can also flatten the density maps and make a “histo-vector” for each exposure (the right panel). Each histo-vector, a one-dimensional distribution, can be considered a fingerprint of the exposure. As can be seen, the four IDs show different patterns (considering the density maps or, equivalently, the histo-vectors). For better content-based matching, each density map here should be compared to Figure \ref{DESOMmap}.

As can be seen, on the histo-vector or the density map, RBT sources have accumulated on a few cells. However, the sources related to the Good image are more uniformly distributed than on the RBT image.  We can realize that the most populated nodes on the RBT density map correspond to those nodes in Figure \ref{DESOMmap} that contain elongated characteristics, e.g., node (19, 2).  So, various patterns can be seen in such plots.  Using the DESOM-1 model, we can obtain all histo-vectors for a set of (new) exposures. Then we will train another deep SOM (DESOM-2) to cluster the histo-vectors.

In summary, for a selected exposure, the output of DESOM-1   can be a histo-vector that contains 625 components.  The sum of a histo-vector will be the total number of detected objects in the associated exposure. For example, if we use $10,000$ massive exposures as the input to DESOM-1, the output will be a matrix with the size (10000, 625). In fact, we convert image information to informative tabular data.  This step is an excellent deep dimensionality reduction step, as well as recognizing useful patterns in each exposure. Such a matrix will be fed to another deep SOM (DESOM-2) to group the 10,000 histo-vectors.

\subsection{DESOM-2}

For each exposure, we can create a histo-vector, which counts the number of thumbnail images that fall within a specific cell on the map (i.e., Figure \ref{DESOMmap}). This method helps us understand the distribution of identified objects and their attributes in the associated exposure. We have already shown some examples of the histo-vectors in the third column in figure \ref{fig:dense}. The x-axis shows the cell index on the DESOM-1, and the y-axis shows the count of those objects identified in the corresponding cell. Exposures that have similar histo-vectors resemble each other. The goal of DESOM-2 is to cluster histo-vector data in order to identify groups of similar images.

To cluster exposures, we could use a sample of histo-vectors with 625 cells as an input to a simple SOM.  However, as mentioned in section \ref{sec:method}, a better solution is to use a DESOM that jointly trains an autoencoder with a SOM. However, since the histo-vectors are not images, instead of using an autoencoder with a CNN architecture we will use a dense autoencoder in the second DESOM (DESOM-2).     

In the lower part of Figure \ref{pipeline}, we show the DESOM-2 structure. The generated histo-vectors are first normalized so that the values within each cell across all the training samples are between $0$ and $1$.  This normalized dataset is then fed to DESOM-2. The autoencoder used in DESOM-2 has a symmetric structure with $2$ hidden layers in the encoder part and $2$ hidden in the decoder. To create a training set for DESOM-2, we use $\sim50k$ exposures, from a catalog, with different characteristics and qualities. We use DESOM-1 and create 50k histo-vectors as a matrix (50k, 625).  The matrix is then used to train DESOM-2. In the model, we choose a map of the size of $20\times20$.  In the following section, we summarize the pipeline used in this work.

\subsection{The pipeline}
\label{sec:pipeline}

\begin{figure*}
\centering
\includegraphics[width=17cm,height=15cm,angle=0]{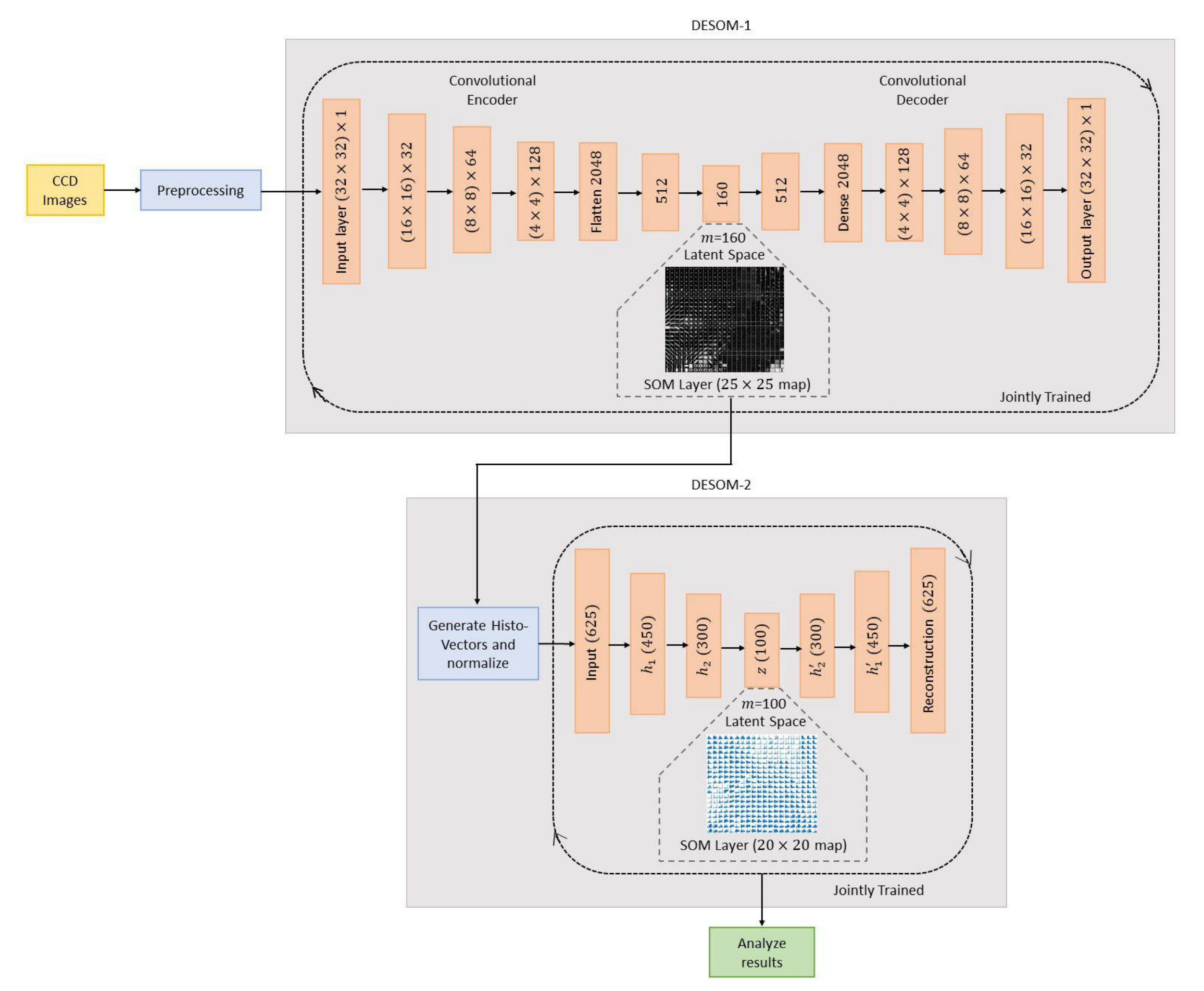}

\caption{The structure of the labeling pipeline proposed in this paper.}
\label{pipeline}
\end{figure*}

Figure \ref{pipeline} shows the pipeline used in this work. First, a pre-processing step is done to prepare normalized cutout images from an exposure (with 36/40 CCDs). Then, the cutout images are fed into the trained model (DESOM-1).  This part of the pipeline produces a density map ($25\times25$) on which the cutout images are distributed. As mentioned before, such a large map provides enough freedom that the algorithm can identify a wide array of objects, imaging conditions, and even noise. Therefore, using DESOM-1, we can get a density map for the exposure. The next step is to flatten the density map and make an associated histo-vector with 625 cells. In other words, each cell in the x-axis of the histo-vector represents a cell on the outcome map, and its associated value shows the number of times an object has been detected in an exposure.

Histo-vectors are then normalized and used in the second part of the pipeline, DESOM-2. The goal of DESOM-2 is to identify groups of histo-vectors that are alike. Since histo-vectors are not images, unlike DESOM-1, the autoencoder part of DESOM-2 is constructed using a fully connected neural network. The output is a $20\times 20$ map which is the final output of this work.  

In summary, an exposure (as the primary input to the pipeline) will land on only one node on the second SOM with a known ID. In other words, an exposure with typically more than 360 million pixels will be mapped to a single point on the second map. Therefore, if we use the $70\rm{k}$ exposures used by T20, we will get a matrix of size (70k, 625) as the output of DESOM-1.  Then we can feed the matrix into DESOM-2.  The $70\rm{k}$ exposures (i.e., their IDs) will then be distributed on the second SOM with $20\times20$ nodes. The IDs within one node should resemble each other. For example, if one of them is labeled as an RBT image, the rest should have the same label. This method can effectively reduce labeling time for substantial databases. We will show some examples in the next section.

\section{Results}
\label{sec:result}
\begin{figure*}
\centering
\includegraphics[width=16cm,height=14cm,angle=0]{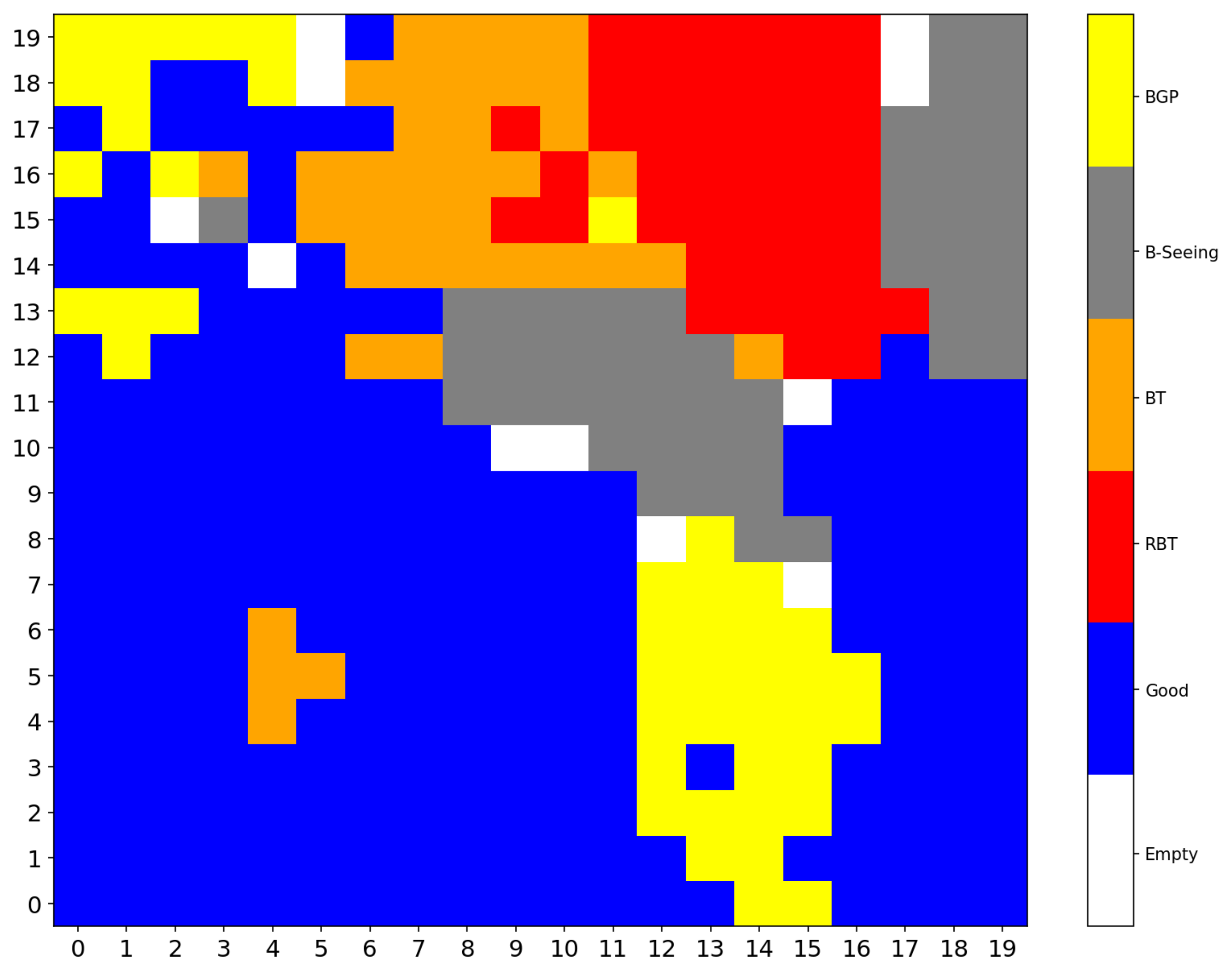}
\caption{A test set of 70k exposures used by T20, which are distributed on DESOM-2. Each exposure in the set has a label of five different classes; Good, RBT, BT, B-Seeing, and BGP.} 
\label{DESOM2}
\end{figure*}

We use the pipeline and apply it to  $\sim70\rm{k}$ IDs (exposures) used by \cite{Teimoorinia20a} as the test set. Each ID in our test set has a label as one of the five different classes (based on five different probabilities); Good, RBT, BT, B-Seeing, and BGP. The pipeline's output is shown in Figure \ref{DESOM2} in which 70k IDs are distributed on the map. Each cell on the map can possess a number of IDs. During the implementation, the IDs within each cell are saved. These IDs, as well as their associated T20 labels, can then be extracted and viewed for evaluation and verification purposes.

\begin{figure*}
\centering
\includegraphics[width=16cm,height=14cm,angle=0]{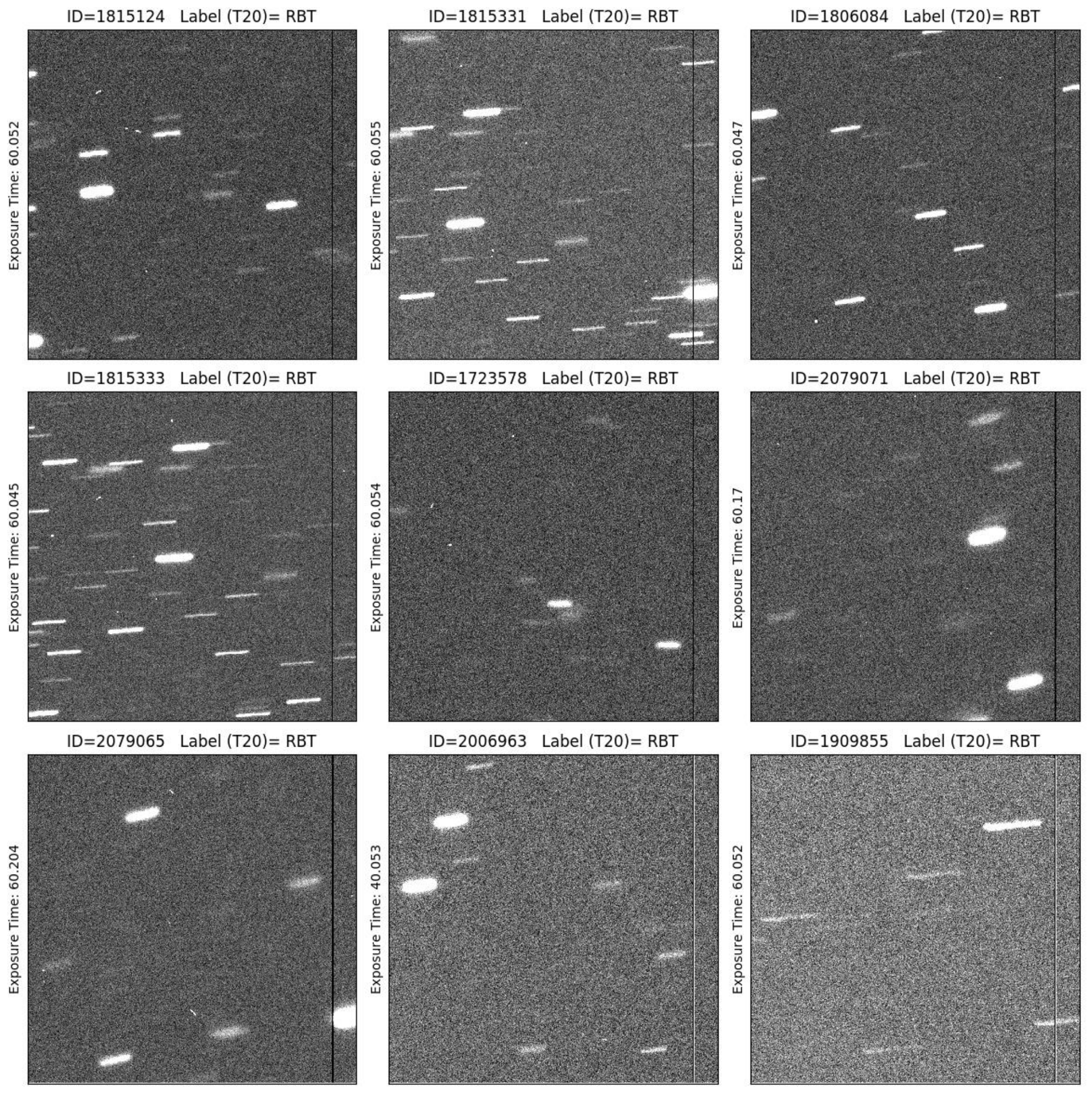}
\caption{Nine randomly selected IDs from cell (15, 19) of Figure \ref{DESOM2}. This example shows that members of this cell have an RBT nature, which is in agreement with T20 results.}
\label{fig-RBT}
\end{figure*}

For example, in Figure \ref{DESOM2} the red cells show those IDs whose RBT probabilities are high, like (Good=$\sim0$, RBT=$\sim1$, BT=$\sim0$, B-Seeing=$\sim0$, BGP=$\sim0$). The blue cells have a high probability of being Good images. The empty cells are cases where the five probabilities are not significantly different from each other, such as (Good$\sim0.2$, RBT=$\sim0.2$, BT=$\sim0.2$, B-Seeing=$\sim0.2$,  BGP=$\sim0.2$). So, we have not labeled them. These empty cells should generally appear at the classes' border, where there could be a transition between two different classes. For example, in Figure \ref{fig-RBT}, we plot nine randomly selected IDs (out of 41) from cell (15, 19).  They are selected from the red cells, which are labeled as RBT IDs by T20.  
It should be noted that, regardless of T20 predictions, we expect to see the same characters for all members of a selected cell. In this example, we show that the selected cell members have an RBT nature. This result agrees with T20, which predicts an RBT nature for the images.

\begin{figure*}
\centering
\includegraphics[width=16cm,height=14cm,angle=0]{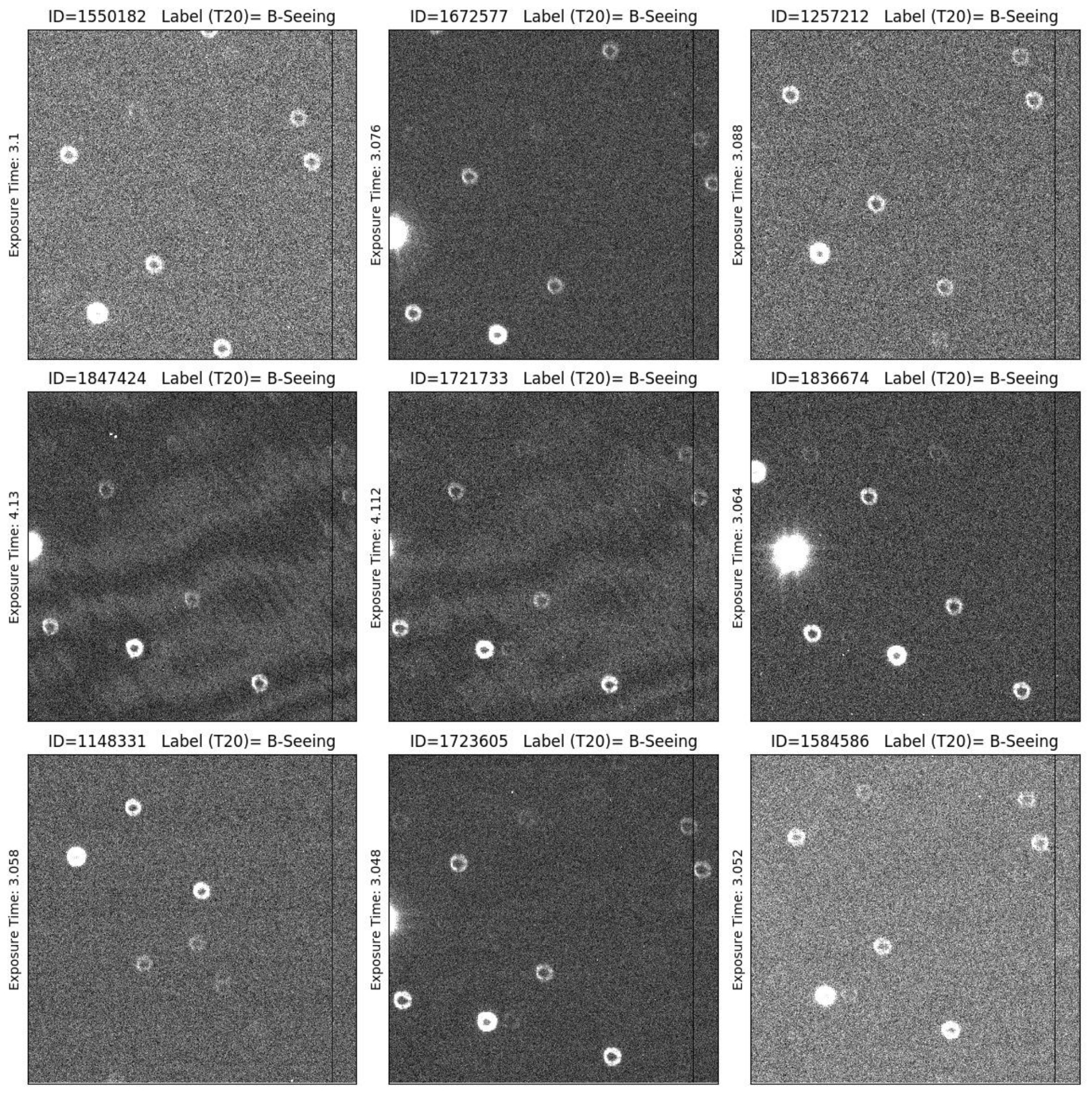}
\caption{Nine randomly selected IDs from cell (19, 19) of Figure \ref{DESOM2}. This example shows that members of this cell have a bad focus nature.}
\label{fig-BF}
\end{figure*}

We show another example of a selected cell's content, cell (19, 19) in Figure \ref{DESOM2}.  The nine randomly selected image IDs (out of 114 in the cell) are shown in figure \ref{fig-BF}. As can be seen, they show the same type. These are images that were taken under the wrong focus condition. The related probabilities in T20 show B-Seeing conditions. T20 used just five classes; therefore, due to this limitation, bad focus images are also classified as B-Seeing (see Figure 12 of T20 and the associated descriptions).

\begin{figure*}
\centering
\includegraphics[width=16cm,height=14cm,angle=0]{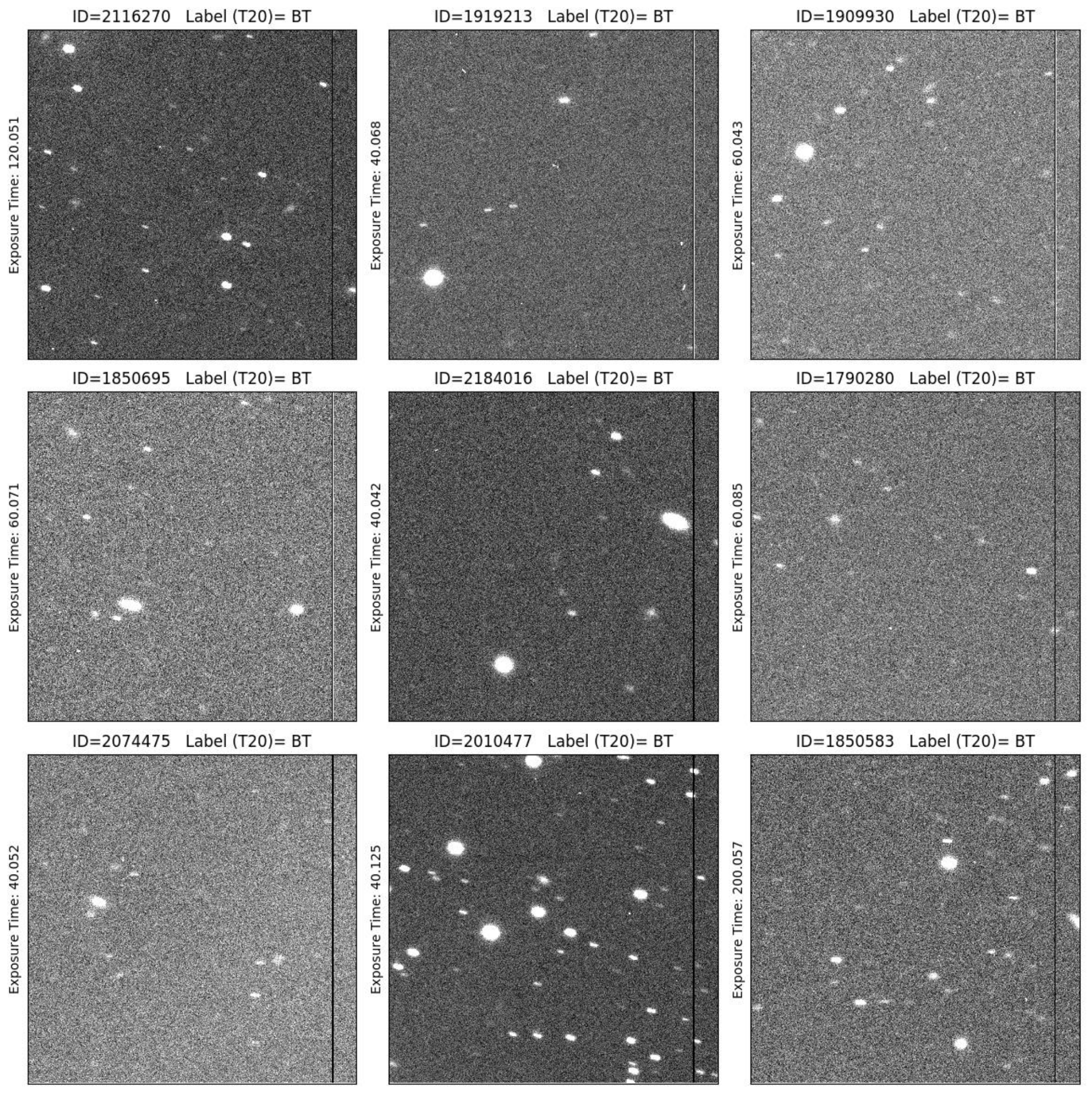}
\caption{Nine randomly selected IDs from cell (8, 19) of Figure \ref{DESOM2}. This example shows that members of this cell have a BT nature.}
\label{fig-BT}
\end{figure*}

\begin{figure*}
\centering
\includegraphics[width=16cm,height=14cm,angle=0]{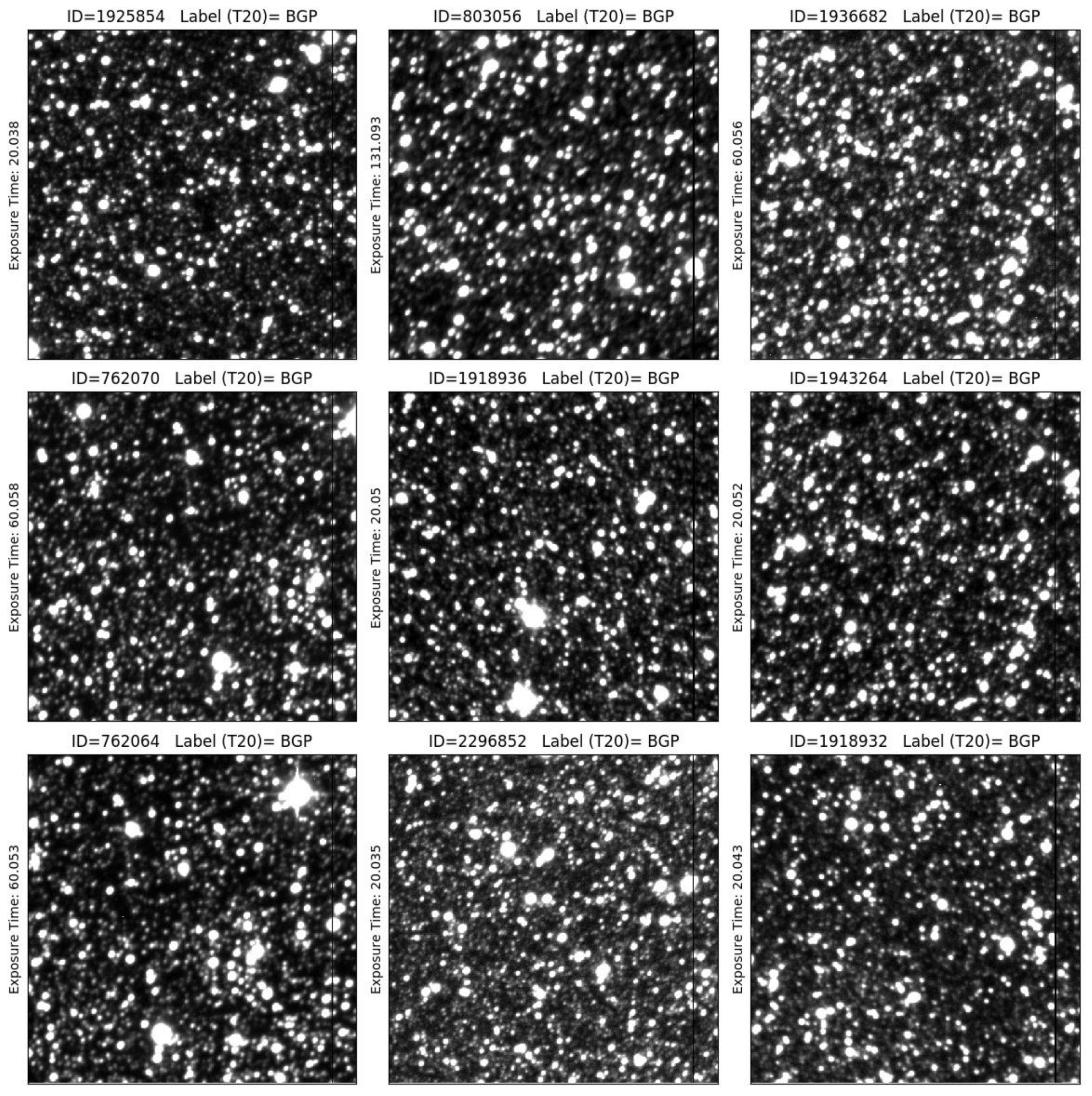}
\caption{Nine randomly selected IDs from cell (0, 19) of Figure \ref{DESOM2}. This example shows that members of this cell have a BGP nature.}
\label{fig-BGP}
\end{figure*}

\begin{figure*}
\centering
\includegraphics[width=16cm,height=14cm,angle=0]{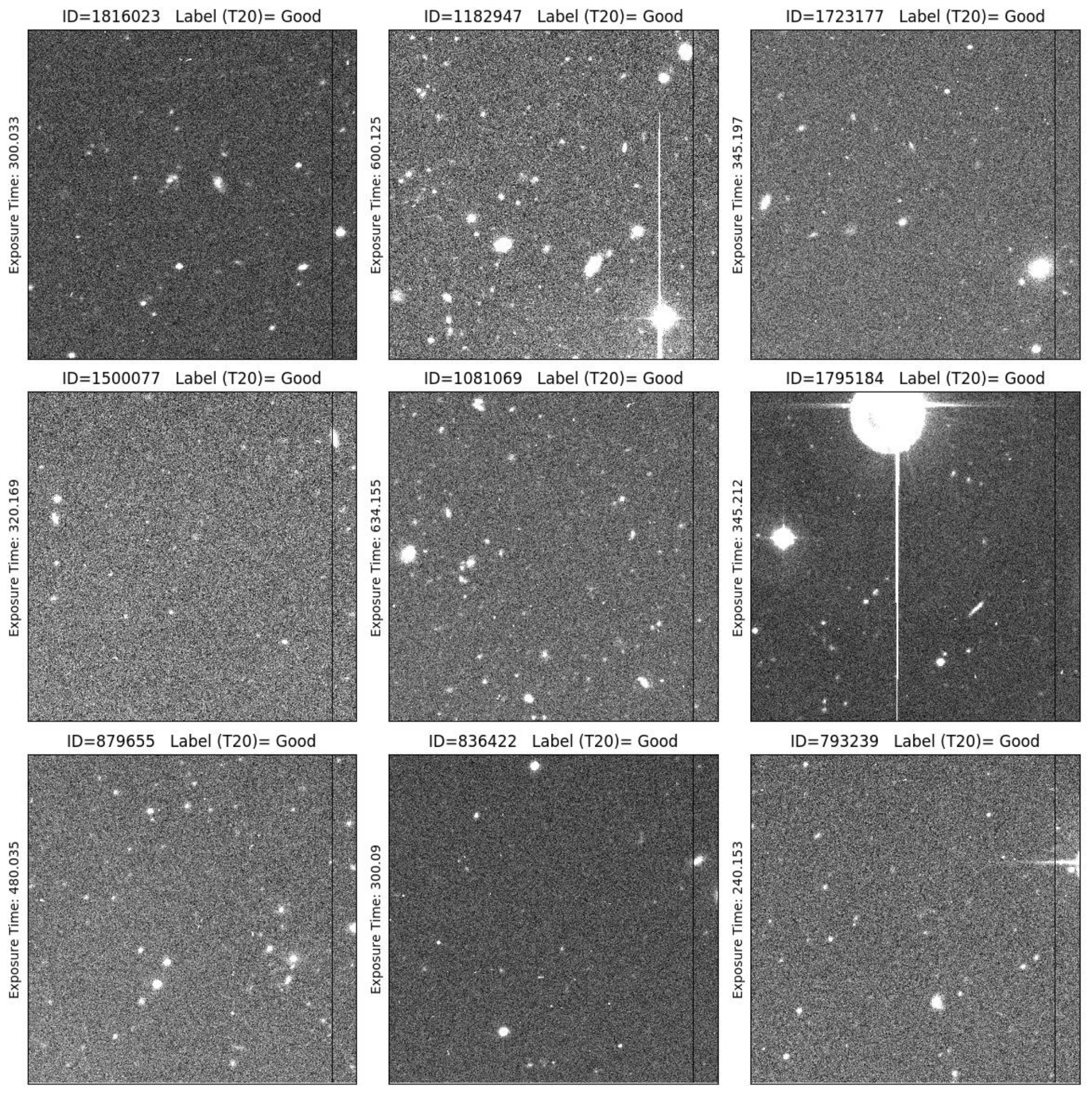}
\caption{Nine randomly selected IDs from cell (1, 0) of Figure \ref{DESOM2}. This example shows that members of this cell have a Good nature.}
\label{fig-GOOD}
\end{figure*}

\begin{figure*}
\centering
\includegraphics[width=16cm,height=14cm,angle=0]{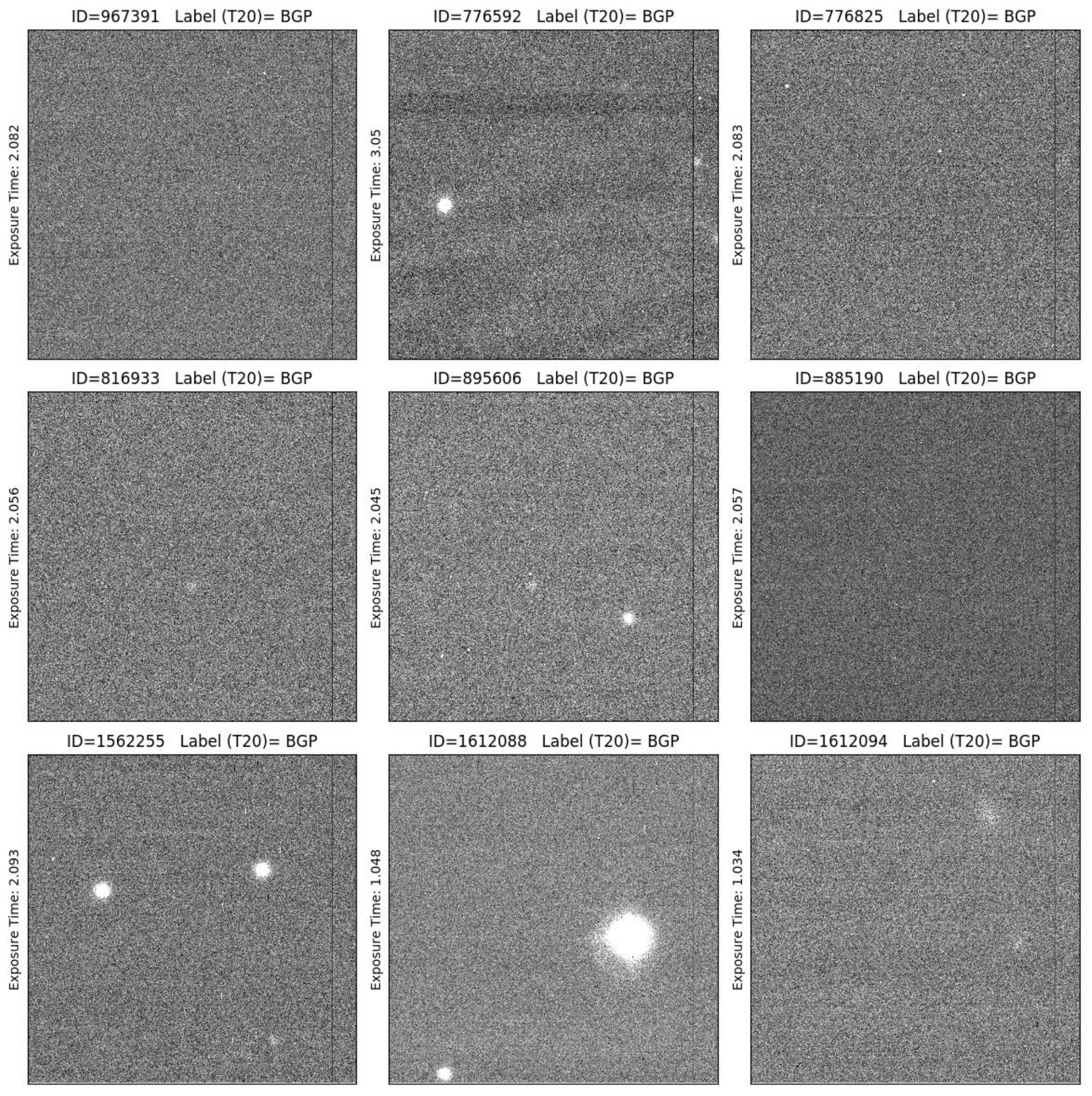}
\caption{Nine randomly selected IDs from cell (14, 7) of Figure \ref{DESOM2}. This example shows that members of this cell show a BGP quality with a low-exposure nature.}
\label{fig-LowExp}
\end{figure*}

\begin{figure*}
\centering
\includegraphics[width=16cm,height=14cm,angle=0]{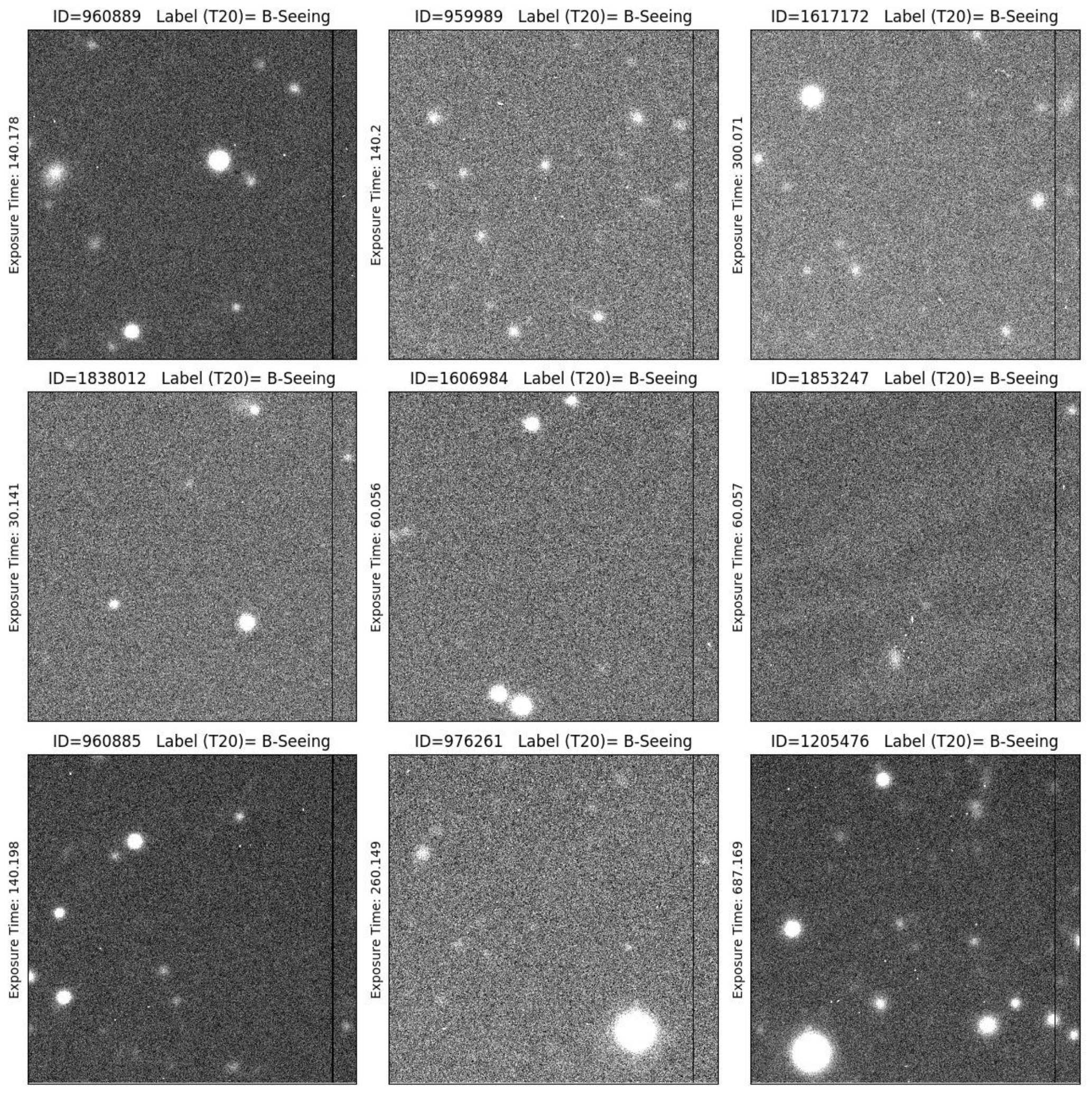}
\caption{Nine randomly selected IDs from cell (11, 11) of Figure \ref{DESOM2}. This example shows that members of this cell have a B-Seeing nature.}
\label{fig-B-Seeing}
\end{figure*}

Figures \ref{fig-BT}-\ref{fig-B-Seeing} shows nine randomly selected images from five different cells.  On top of the images, the labels predicted by T20 are shown. There is good agreement between T20 labels and image quality. However, again, it is important to see that each group's content shows the same nature. For example, in Figure \ref{fig-BGP} the nine images are very similar, even though they are taken from different IDs.  They are very crowded images that are classified as BGP in T20.  Many of the objects in the images have a high potential to be influenced by neighboring contaminant objects.  However, these images might be labeled as usable images in some scientific contexts. Therefore, if there are half a million different IDs, taken in different quality and with different types, an expert needs to explore only some cell members instead of the half-million IDs. The pipeline (as a jupyter notebook), the models’ details and all members of the 400 cells (for our test set) can be found here:~\href{https://github.com/shishehchi/AstroUnsupervisedDeepClustering}{(pipeline-link)}.

\section{Discussion}

\label{sec:discussion}
This paper aims to employ useful information taken from a MegaCam image (as the input to our pipeline) and assign each image ID to a cell on DESOm-2. Similar images occupy one place on the map. In this way, we can cluster different images and then label them in a very efficient way. In the following sections, we compare our results with T20 results and describe some technical details and potential applications.

\subsection{Comparing unsupervised and supervised results}

In Figure \ref{DESOM2}, we have utilized the $\sim70k$ exposures used by T20 as a test set.  T20 apply the set in a supervised mode and, using five labels, classify the images into five different classes. As mentioned in Sec. \ref{introduction}, supervised methods are potent approaches that excel at optimizing performance in a well-defined task. However, they cannot generalize beyond the assigned items they are trained on.  One of the restrictions in T20 is the limited number of classes. For example, in Figure \ref{DESOM2}, there are two separated  B-Seeing regions on the map. As shown in Figure \ref{fig-BF} and \ref{fig-B-Seeing}, they have different properties discovered by this work's unsupervised methods—an advantage of using data-driven approaches.

\begin{figure*}
\centering
\includegraphics[width=16cm,height=14cm,angle=0]{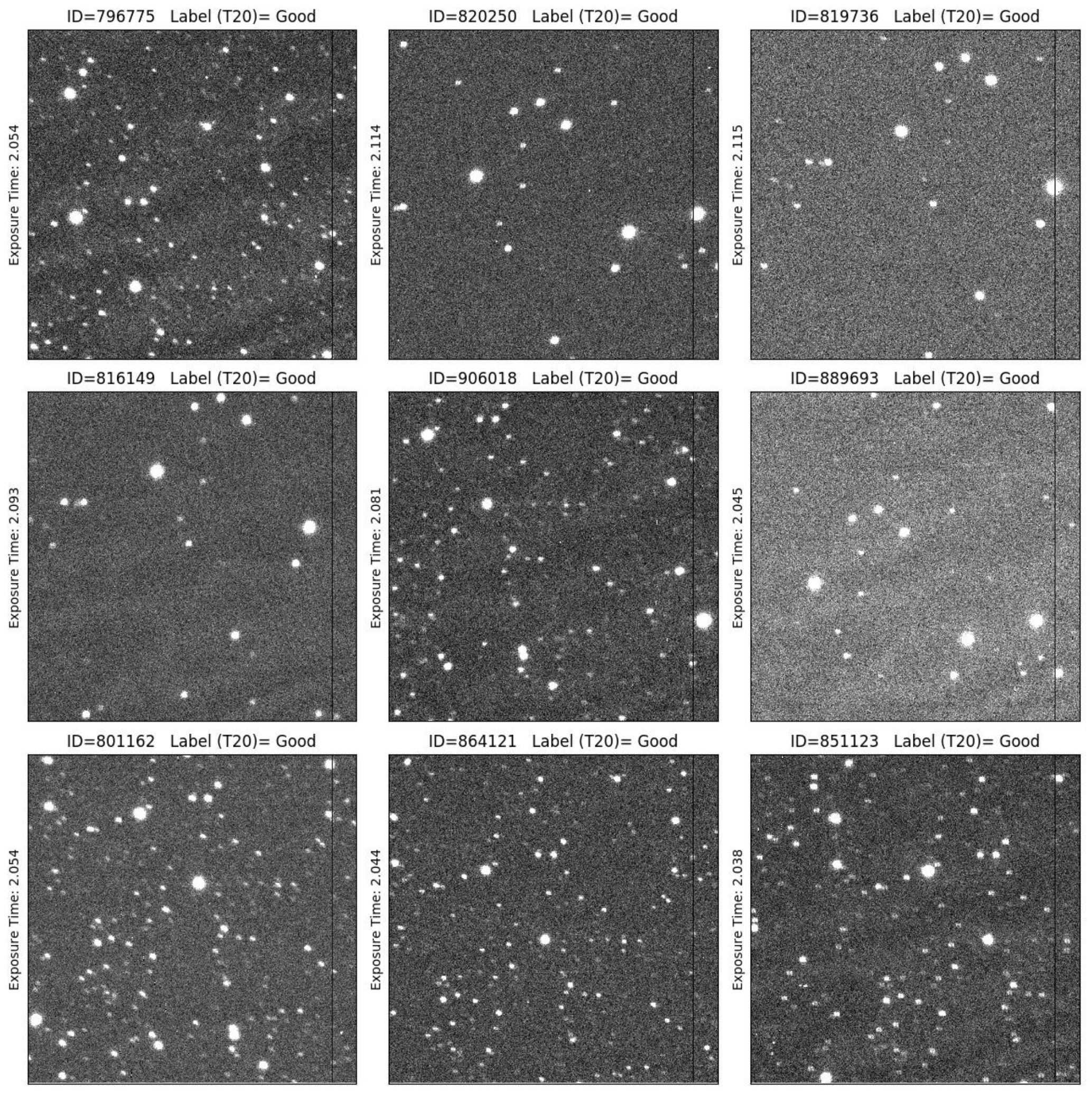}
\caption{Nine randomly selected IDs from cell (13, 3) of Figure \ref{DESOM2}. This example shows that members of this cell have a Good nature (based on T20); however, they are located in the middle of the BGP/low-exposure region on the map and may not be considered Good images according to our unsupervised method.}
\label{fig-Good-LowExp}
\end{figure*}

Another example is related to BGP cells on the map (i.e., the yellow regions). Notice that there are many BGP areas that are far apart from each other.  As can be seen in Figure \ref{fig-BGP} and \ref{fig-LowExp}, very crowded images (with background problems) and low-exposure ones (another type of background problems) are classified into one class (i.e., BGP) by T20.  The unsupervised method automatically separates them into two different regions, because the IDs in that class have more freedom to move among 400 cells in order to find more suitable groups. Remarkably, we do not use any information on image exposure times in our pipeline; however, low-exposure images are grouped together. This grouping could be due to more fluctuations that might exist in the low-exposure and noisy images.  An interesting example in this respect is linked to the blue cell (13, 3) surrounded by BGP cells (i.e., yellow, low-exposure cells). The blue cell members are labeled by T20 (with a relativity lower probability) as Good images. We show some members of this blue cell in Figure \ref{fig-Good-LowExp}. As can be seen, they are all low-exposure images. An expert might not label them as Good images in the unsupervised methods. Another example of low-exposure images is seen in Figure \ref{fig-BF}. They are bad focus images; however, all have low exposure times as well.

The supervised and unsupervised methods are in good agreement in most cases. There are some disagreements because of the limitations of the supervised method. The unsupervised method shows more flexibility and freedom, and an expert can use this facility to label images in a more effective and informative way.  Once the expert investigates some of the cells in DESOm-2, the pipeline is ready to label any new MegaCam image.

So far, we have evaluated our pipeline qualitatively and compared it with the supervised method. To calculate quantitative performance metrics such as accuracy, precision or recall (Eq.~\ref{eq-metrics}), reliable ground truth is needed. In Eq.~\ref{eq-metrics}, TP (True Positive) is the number of Good images that are correctly predicted. TN (True Negative)  is the number of (all) Bad occurrences that are correctly predicted.  FN (False Negative) indicates the number of instances that belong to the Good group but are labeled as Bad. FP (False Positive) shows the number of examples that belong to the Bad group but are predicted as Good.  In this project, we can consider two sources of information as ground truth: T20 labels or expert knowledge. The former is already provided to us; however, we have shown and discussed a number of cases where the supervised method did not provide an accurate label. As an example, it is shown in figure \ref{fig-Good-LowExp} where images with BGP nature are labeled as Good by T20. However, the unsupervised method successfully grouped these images as similar images in one cell.

Furthermore, the supervised method is only limited to five labels, making it incompatible with being compared with an unsupervised output map of 400 degrees of freedom. Despite these fundamental challenges, we calculated the quantitative performance metrics assuming that the supervised labels are the true labels shown in figure \ref{DESOM2}. A more reasonable approach can treat it as a binary classifier where RBT, BT, B-Seeing and BGP are considered bad labels. By iterating through the IDs in each cell, ignoring the white cells, and taking into account the label provided by T20 for each ID, it is possible then to calculate the metrics. Accuracy, precision and recall  are $0.940$, $0.946$ and $0.978$, respectively, which shows a good agreement between the two approaches.

\begin{equation}
\rm{Accuracy=\frac{TP +TN}{FN+FP+TP+TN}}, ~~ 
\rm{Precision=\frac{TP}{TP+FP}},  ~~
\rm{Recall=\frac{TP}{TP+FN}} 
\label{eq-metrics}
\end{equation}

Alternatively, we can consider expert knowledge as the true labels for each image. In this case, an expert has to examine the IDs within one cell to determine whether they are Good or Bad images. Mismatches between expert knowledge and the unsupervised predictions in a cell can be considered false positive or false negative depending on the majority class within the cell. Assigning quality labels to all the images used in our pipeline by manually evaluating them is impossible. Instead, we have randomly selected 2000 IDs from 40 randomly selected cells (50 IDs per cell), and by inspecting sample images manually, we have assigned a Good or Bad label to each image. The number of Good and Bad labels is almost equal; therefore, we have a balanced sample. The label of the majority class within a cell is set as the label of the cell. Accuracy, precision and recall, in this case, are $0.958$, $0.929$ and $0.991$, respectively. A high value of precision shows an algorithm's ability to decrease the number of exposures incorrectly classified as FP. At the same time, recall represents the algorithm's capacity to minimize the number of exposures wrongly identified as FN. A high value of recall here indicates a `negligible' number of Good IDs labeled as Bad. Here, precision is relatively lower than recall, indicating a higher number of FP compared to FN. Most FP cases here are related to those IDs with background problems where the objects alone appear good; however, the background does not seem reasonable.

\subsection{Some technical details}

Some technical points could be useful for people who want to use the method or make a similar pipeline. As we already mentioned, a source- or signal-detector software package is used to cut detected sources out of the IDs. The number of cutout images from some IDs (especially crowded images such as Figure \ref{fig-BGP}) could be too high. So, storing and pre-processing them to create associated histo-vectors could be costly. One way to overcome this issue is to randomly choose a “reasonable” number of detected sources and cut them out of the corresponding CCD for the pre-processing step. To create the histo-vectors, we need the total number of detected sources rather than that of the sample. Therefore, when selecting a sample, at the same time, we can record the number of total detected sources as a weight factor. When we build the associated histo-vector for an ID, we can weigh it by the recorded number.

Another important point is that, there are two temperature parameters Tmax and Tmin in DESOM models as described in the paper. Tmin is an important   parameter  that must be set when defining the model (i.e., a hyper-parameter). If we use a very small value for Tmin, we convert the SOM to a Kmean method. In this case, the neighborhood kernel functionality in SOM will be useless, and as a result, we would see more separated and divided regions in DESOM.  Some prior information from the training set could help choose a suitable value for Tmin. For example, we could use just two distinct groups, such as RBT and B-Seeing, as a training set and observe the cells' connectivity on DESOM-2. Hence, we may find an optimum value for Tmin. 

\subsection{Some possible applications}

One of the labeling systems' applications could be to use the predicted labels as a target for a supervised method to provide probabilistic results. In the following, we mention some potential and more specific applications for our method.

One application of DESOM-1 would be to measure the prototypes' different geometrical parameters, such as elongation or ISO-AREA \citep[see, e.g.,][]{bertin96}, and save it as a fixed database. So, for an ID whose cutout images are ready to be distributed on DESOM-1, we may present statistical information of detected objects' geometrical properties in the ID.

Some applications, such as preparing training sets for some classification uses, or separating merger galaxies from non-mergers, need non-deficit cutout images. For such applications, we could train a relevant DESOM-1 using a sample of available cutout images. Then we can use the trained model. The cutout images could first be filtered out using the trained DESOM-1, and we may eliminate images such as those we can see in the cell (1, 12) in Figure \ref{DESOMmap}. 

In the data center, the images' content can be investigated by a trained DESOM such as Figure \ref{DESOMmap}. For  example,  a density map of a CCD image that shows crowded cells around cells (2, 2) can be marked before being downloaded by the database users. In other words, DESOM-1 may be used as a content-based recommendation system.

\section{Conclusion}
\label{sec:conclusion}

We present a method that combines two advanced deep learning algorithms to cluster and assess astronomical images' quality in a fully unsupervised mode. The data used in this project is a set of selected images from the MegaCam instrument mounted on the Canada-France-Hawaii Telescope. Each exposure (ID) contains 36 or 40 CCDs, where each CCD has a size of $2048\times4612$ pixels resulting in $\sim350$ million pixels per exposure. The first part of the pipeline, DESOM-1, comprises a CNN autoencoder jointly trained with a SOM. An extensive collection of different sources, as small cutout images, are fed to DESOM-1 to create 625 prototypes. This step ensures that we have a `complete' set of the prototypes (with different characteristics and nature) representing Megcam's sources.  The output of DESOM-1 is a $25\times25$ map of prototype vectors representing the objects in the training data set. After the pipeline has successfully identified various shapes, rotations, sizes and imaging conditions present in the training data set, we use the trained DESOM-1  to generate an output for a given exposure called a histo-vector.  We use a significant sample of IDs and generate the corresponding histo-vectors. The next step is to cluster similar histo-vectors by using the DESOM-2 model, a dense autoencoder jointly trained with a SOM. The result is a $20\times20$ map. Each cell on the map represents groups of exposures that are similar to each other.
In summary, an exposure with more than 350 million pixels will be mapped to only one cell on the second map. We have concluded that our unsupervised method successfully identifies various imaging qualities, and groups similar images together in one cell. To assess the system's ability and performance, we compared our unsupervised method with the supervised method presented by \cite{Teimoorinia20a}. The unsupervised and supervised methods show a good agreement. However, we observed that while the supervised method only provided five classes of image qualities, our unsupervised method automatically identified new classes in the data set that were completely unknown in the supervised method. A manual inspection of unsupervised outputs shows a very good clustering result with accuracy, precision and recall of $0.958$, $0.929$ and $0.991$, respectively.


\bibliography{main}{}
\bibliographystyle{aasjournal}



\end{document}